\documentclass[aps,prx,twocolumn,superscriptaddress,floatfix]{revtex4-1}
\usepackage{amsfonts,hyperref}

\usepackage{graphicx}
\usepackage{amsmath}
\usepackage{amssymb}
\usepackage{bm}% bold math
\usepackage{gensymb}
\usepackage{braket}

\newcommand{\tr}{\mathrm{tr}}
\newcommand{\xid}{\xi_D}
\newcommand{\xic}{\xi_\chi}

\begin{document}
\title{Finite correlation length scaling with infinite projected entangled-pair states}

\author{Philippe Corboz}
\affiliation{Institute for Theoretical Physics and Delta Institute for Theoretical Physics, University of Amsterdam, Science Park 904, 1098 XH Amsterdam, The Netherlands}

\author{Piotr Czarnik}
\affiliation{Institute of Nuclear Physics, Polish Academy of Sciences, Radzikowskiego 152, PL-31342 Krak\'ow, Poland}

\author{Geert Kapteijns}
\affiliation{Institute for Theoretical Physics and Delta Institute for Theoretical Physics, University of Amsterdam, Science Park 904, 1098 XH Amsterdam, The Netherlands}

\author{Luca Tagliacozzo}
\affiliation{Department of Physics and SUPA, University of Strathclyde, Glasgow G4 0NG, United Kingdom}
\date{\today}

\begin{abstract}
We show how to accurately study  2D quantum critical phenomena using infinite projected entangled-pair states (iPEPS). We identify the presence of a finite correlation length in the optimal iPEPS approximation to Lorentz-invariant critical states which we use to perform a finite correlation-length scaling (FCLS) analysis to determine critical exponents. This is analogous to the one-dimensional (1D) finite entanglement scaling  with  infinite matrix product states. We provide  arguments why this approach is also valid in 2D by identifying a class of states that despite obeying the area law of entanglement  seems hard to describe with iPEPS. We apply these ideas to interacting spinless fermions on a honeycomb lattice and obtain critical exponents which are in agreement with Quantum Monte Carlo results. Furthermore, we introduce a new scheme to locate the critical point without the need of computing higher order moments of the order parameter. Finally, we also show how to obtain an improved estimate of the order parameter in gapless systems, with the 2D Heisenberg model as an example. 
\end{abstract}

%71.10.Fd	:	Lattice fermion models (Hubbard model, etc.)
%71.27.+a	Strongly correlated electron systems; heavy fermions
%71.10.Hf	Non-Fermi-liquid ground states, electron phase diagrams and phase transitions in model system
%02.70.-c		:	Computational techniques; simulations

%02.70.-c		:	Computational techniques; simulations
%71.10.Fd	:	Lattice fermion models (Hubbard model, etc.)
%03.67.-a		:	Quantum information
%67.85.-d,  ultracold Gases
%05.30.Fk, quantum statistical mechanics
%75.10.Jm Heisenberg model
%75.10.Jm 	Quantized spin models, including quantum spin frustration
%75.10.Kt 	Quantum spin liquids, valence bond phases and related phenomena
%75.40.Cx 	Static properties (order parameter, static susceptibility, heat capacities, critical exponents, etc.)

\maketitle

%In this paper we show the existence of a similar phenomenon to the finite entanglement scaling of 1D quantum many-body systems in the context of  2D quantum many-body systems.

In recent years there has been a very active development of tensor-network variational ansatzes for describing strongly correlated quantum-many-body systems~\cite{Verstraete08, schollwoeck2011,orus_practical_2014, bridgeman_hand-waving_2017, haegeman_diagonalizing_2017}. 
Tensor networks exploit the fact that ground states and low-energy states of local Hamiltonians are typically only weakly entangled, where the entanglement entropy of a region scales only with its surface rather than with its volume - a property known as the area-law of entanglement  \cite{hastings_area_2007,wolf_area_2008,eisert_area_2008,amico_entanglement_2008,masanes_area_2009,laflorencie_quantum_2016}. 
As a consequence, the information contained in these states can be compressed and described by using specific tensor networks where the elementary tensors have finite bond dimension $D$, such as  one dimensional matrix-product states (MPS)~\cite{affleck_rigorous_1987,ostlund_thermodynamic_1995,perez-garcia_matrix_2007} and  2D projected entangled-pair states (PEPS)~\cite{verstraete2004,Verstraete08,jordan2008} (also called tensor product states~\cite{nishino01,nishio2004}).

In particular, in 1D, infinite MPS (iMPS) have successfully been used to characterize the universal properties of critical systems. This at first sight seems counter-intuitive since 1D critical states violate the area-law of entanglement \cite{callan_geometric_1994,calabrese_entanglement_2007,vidal_entanglement_2003} while an iMPS with finite $D$ can only describe states fulfilling the area-law, i.e. gapped states with a finite  correlation length $\xi$ \cite{affleck_rigorous_1987,ostlund_thermodynamic_1995,perez-garcia_matrix_2007}.
However, with increasing bond dimension $\xi_D$ increases, improving the accuracy of the approximate state in a systematic way. As a result local observables follow the  universal scaling laws characteristic of the underlying critical point. In the scaling regime $\xi_D$ acts as a cutoff on the exact, diverging  correlation length, similarly to a finite system size. The possibility to tune $\xi_D$ by varying the bond dimension can be practically used to extract  critical exponents in a very similar way as in standard finite-size scaling approaches. This powerful approach is known as finite entanglement scaling or also called finite correlation-length scaling (FCLS)~\cite{tagliacozzo08,pollmann2009,pirvu12}.

An important question is whether a similar approach can also be designed in 2D, which would be highly desirable since most critical exponents can only be computed numerically \footnote{See however also the recent developments of the analytical bootstrap \cite{rychkov_epsilon_2015, gliozzi_generalized_2017}.}.
However, the situation in 2D seems different because unlike in 1D there exist critical states with an area law~\cite{liao_heisenberg_2016,liao_gapless_2017,poilblanc_quantum_2017} and there are known examples of exact critical iPEPS with a finite $D$~ \cite{verstraete_criticality_2006}. The latter  include 
2D classical states \footnote{e.g. the 2D classical partition function of the critical 2D Ising model represented as an iPEPS with $D=2$.} and ground states of generalized  Rokhsar-Kivelson (RK) Hamiltonians at their critical point. When the RK Hamiltonian is critical the low energy excitations' energy-momentum dispersion relation is $E(k) \propto k^z$ with $z\ge 2$ and the RK states effectively describe the partition function of  2D classical models ~\cite{henley_classical_2004,ardonne_topological_2004,castelnovo_quantum_2005,isakov_dynamics_2011,tagliacozzo_tensor_2014,zohar_fermionic_2015,zohar_projected_2016,poilblanc_quantum_2017}. 
Beside RK states, it is currently still unclear whether a generic quantum critical 2D state, obeying the area law, can be exactly represented by finite-$D$ iPEPS.

Here we will focus on a special case of quantum phase transition,   a critical point with  low-energy excitations exhibiting a linear energy-momentum dispersion relation, $E(k) \propto k$. The linear dispersion is the footprint of an enhanced  emerging symmetry, where energy and momentum (or space and time) play a very similar role and thus these critical points are called Lorentz-invariant critical points.  For such critical point  there are no known examples of a finite-$D$ iPEPS that exactly represents the critical state.

In this paper we provide arguments justifying that an iPEPS with finite $D$  can not, in general, represent a Lorentz invariant critical state exactly. The finite $D$ always induces a finite correlation length $\xi_D$ in the iPEPS state, in complete analogy to the 1D case.  Lorentz invariant critical points could thus describe a class of states which, despite fulfilling the area law of  entanglement, cannot be faithfully represented by an iPEPS with a finite $D$ (see also \cite{ge_area_2016}). As a positive consequence, this allows us to the apply the ideas of FCLS also in 2D for the accurate and systematic study of quantum critical phenomena. 

In order to demonstrate the applicability and power of FCLS in 2D we present state-of-the-art simulation results for interacting spinless fermions on the honeycomb lattice where we find critical exponents in agreement with quantum Monte Carlo results. Furthermore, we introduce a new scheme to locate the critical point, based on the order parameter $m$ and its derivative (called $m'/m$-approach). Similar to the usual Binder cumulant approach, this scheme does not require the a-priori knowledge of the critical exponents, but it is simpler since it is not based on higher moments of the the order parameter, which  are  computationally expensive to obtain with iPEPS. We also show how FCLS can be used to obtained improved estimates of order parameters in gapless systems, with the 2D Heisenberg model as an example.

\begin{figure}[h]
\includegraphics[width=1\columnwidth]{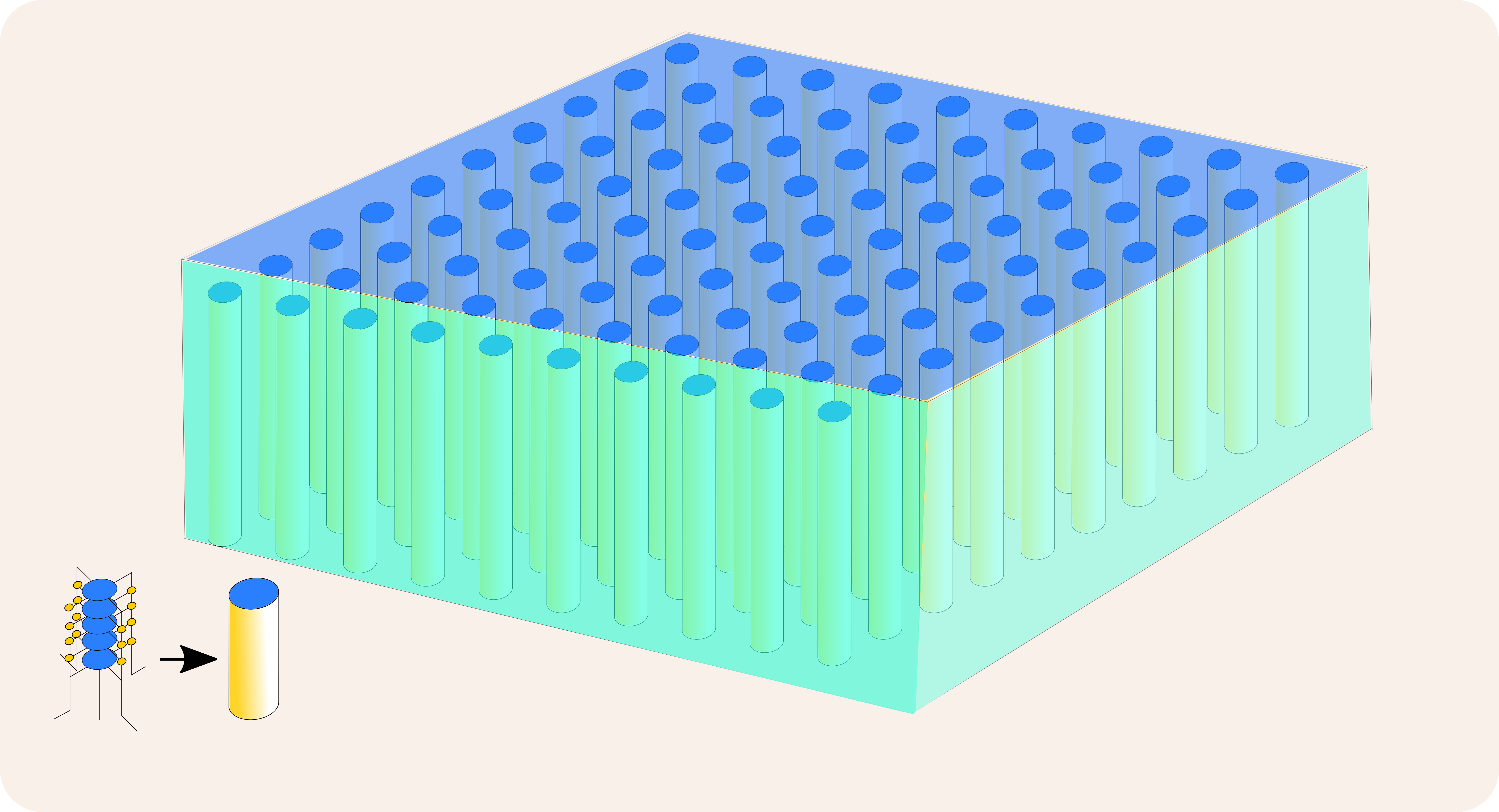}
\caption{(Color online) In the scaling regime, close to a Lorentz invariant critical point, the low energy physics of a 2D quantum lattice system is described by a continuous 3D field theory. In the same regime, the iPEPS wave function describes a more complicated landscape, where the 3D space is pierced by infinitely long valleys separated by distances of the order of  the lattice spacing. 
The correlation length along those towers provides the infra-red cut-off that is responsible for the appearance of a gap in the system, which we observe in our numerical simulations.  
These towers are generated by the finite bond dimension in iPEPS (shown in the lower left corner of the illustration), as explained in detail in the main text.
\label{fig:towers}}
\end{figure}

Our reasoning seems to indicate that the mismatch between finitely correlated iPEPS  and critical states has a geometric origin. The finite $D$ in iPEPS,  in the scaling regime, transforms the  continuous space-time into a landscape of towers in imaginary time separated by valleys. The inter-tower separations are at the scale of the lattice spacing, but the finite correlation along the towers provides the infra-red cutoff that is ultimately responsible of the appearance of the finite correlation length in the system, as sketched in Fig. \ref{fig:towers}.

This paper is organized as follows: the next section provides a brief introduction to iPEPS, followed by a heuristic discussion about the effects of encoding the ground state of a 2D Hamiltonian at a Lorentz invariant quantum critical point with an iPEPS in Sec.~\ref{sec:theory}. In Sec.~\ref{sec:fes} we explain how to perform a FCLS analysis with iPEPS in practice. In Sec.~\ref{sec:results} we present our numerical results for the interacting spinless fermion model and explain the $m'/m$ scaling approach. The extrapolation technique based on FCLS to improve the extrapolation for the order parameter to its exact infinite $D$ value is explained in Sec.~\ref{sec:op}. Finally, in Sec.~\ref{sec:conclusion} we summarize our main findings and conclusions.

\section{Infinite projected entangled-pair states}
\label{sec:ipeps}
An iPEPS~\cite{verstraete2004,Verstraete08,jordan2008} (also called tensor product state~\cite{nishino01,nishio2004}) is an efficient variational tensor network ansatz for two-dimensional ground states of local Hamiltonians in the thermodynamic limit, and can be seen as a natural generalization of (infinite) matrix product states  to two dimensions. The ansatz consists of a periodically repeated unit cell of tensors with one tensor per lattice site. In the present work we used a cell with 2 tensors arranged in a checkerboard pattern. Each tensor has one physical index, carrying the local Hilbert space of a lattice site, and 4 (3) auxiliary indices on a square (honeycomb) lattice. As in an MPS the accuracy of the ansatz is systematically controlled by the bond dimension $D$ of the auxiliary indices. In recent years iPEPS has become a very powerful approach which has been applied to a broad range of challenging problems, including frustrated spin systems and strongly correlated electron systems, see e.g. Refs~\cite{Zhao12,  Corboz12_su4, xie14, Corboz13_shastry, gu2013, corboz14_tJ, picot15a, picot15, liao16, niesen17, zheng17, poilblanc_quantum_2017, haghshenas17} and references therein. 

The optimization of the tensors (i.e. finding the optimal variational parameters) in this work is done by a combination of imaginary time evolution algorithm~\cite{corboz2010,phien15} and variational optimization~\cite{corboz16b} (see also Ref.~\cite{vanderstraeten16}). The contraction of the 2D tensor network is done using a variant~\cite{corboz14_tJ}  of the corner-transfer matrix (CTM) method~\cite{nishino1996, orus2009-1}, where the accuracy of the contraction is controlled by dimension $\chi$ of the boundary tensors, see e.g. Ref.~\cite{corboz2010} for details. In order to improve the efficiency we exploited global Abelian symmetries in the tensor network ansatz~\cite{singh2010,bauer2011}.

\section{Lorentz invariant critical points} \label{sec:theory}

Lorentz-invariant critical points are points in the phase diagram of a many-body systems in which i) the Hamiltonian is gapless and ii) the low-energy excitations'  dispersion relation depends linearly on the momentum $k$, $E(k)=v k$, where $E$ is the energy of the excitation and  $v$ is the sound velocity.

The linear  dispersion relation implies that the system  at low energy becomes Lorentz invariant. The Lorentz transformations mix space and time. In order to understand the role of ``time'' in the ground state of a critical system it is better to appeal to the universality of the critical point. In this way we can describe the same scenario in terms of a classical 3D system where the extra dimension will represent the ``time''. The low-energy emerging properties at the critical point of the both the 2D quantum system and the 3D classical system are the same, provided that we choose a critical 3D classical system in the same universality class as the 2D quantum system.

One way to construct a 3D classical system in the same universality class of our 2D quantum system is through the correspondence between classical and quantum mechanics. From the partition function of a classical model we can write the ground state space projector $\ket{\Omega}\bra{\Omega}=\lim_{\beta \to \infty} \exp( -\beta H)/Z$ with $Z=\textrm{tr}( \exp( -\beta H))$, \footnote{If $H$ is gapless, a projector onto a well defined state can be obtained by first opening an infinitesimally small gap of order $\epsilon$ modifying  $H \to H_{\epsilon}$, and then taking the limit $\beta \to \infty$. We will avoid similar details in the rest of the section.} and the ground state expectation values $\langle {\cal{O}} \rangle  = \tr { \left(\ket{\Omega}\bra{\Omega}{\cal{O}} \right)}$ with $\cal{O}$ an arbitrary operator. This is done by i) dividing the euclidean-time $\beta$ into many small intervals $\beta/N_t =\delta \beta$. In this way ii) the expectation value of an operator is written in terms of a large power of a certain operator $\langle {\cal{O}} \rangle =  \tr{\left(T^{N_t}\cal{O} \right)}$ where $T=\exp( -\delta \beta H)$. We iii) insert the resolution of the identity in a preferred basis of the Hilbert space $\mathbb{I}=\sum_{\set{n}}\ket{\set{n}}\bra{\set{n}}$ before and after each $T$. This allows to identify a ``classical'' transfer matrix. iv) As a consequence of the locality of the Hamiltonian $H$ and the fact that  $\delta \beta \ll 1$, $T$ can be expressed approximately (with an error that scales to zero as $\delta \beta \to 0$ faster than $\delta \beta ^2$) as the contraction of local Boltzmann weights, by for example performing a Trotter-Suzuki expansion of $T$ together with a singular value decomposition of the resulting  local  terms \cite{vidal_classical_2007,vidal_efficient_2004,orus_infinite_2008}. 
As a result we can express the ground space projector  of a  2D Hamiltonian (and consequently all ground state expectation values)  as  the infinite contraction of a simple 3D tensor network.

The construction holds for any 2D quantum system and becomes exact in the limit  $\delta \beta \to 0$. The emerging Lorentz symmetry close to a Lorentz invariant critical point allows to exchange the role of the euclidean-time ($\beta$) with the one of any of the space directions. As a consequence, the correlation length in the time direction is proportional to the correlation in the space direction $\xi_s=v \xi_t$, where $v$ is the velocity appearing in the low-energy dispersion relation and $s$ and $t$ denote the space and time correlation length and time. 

We now give an argument in favor of the fact that by encoding the 3D tensor network in a 2D iPEPS with finite bond dimension we force  the correlation time $\xi_t$ to be finite. 

\subsection{Finite correlation time}
\begin{figure}[h]
\includegraphics[width=1\columnwidth]{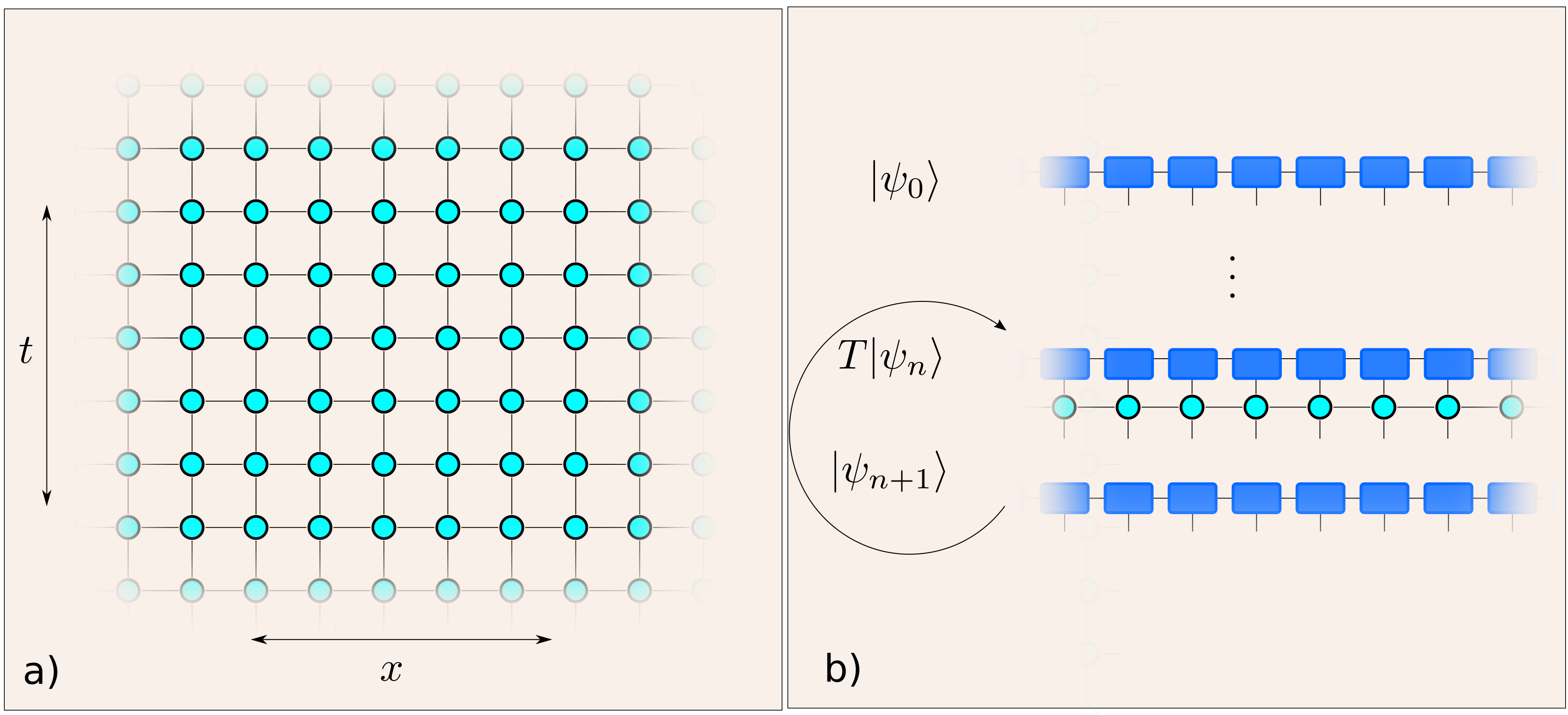}
\caption{(Color online) a) The ground space projector can be represented as an infinite 3D tensor network, spanning the three direction $x,y,t$. As usual tensor are represented here by geometric shapes and lines attached to them are their indices. A line joining two tensors represents the tensor contraction.  Here we represent a projection of the 3D tensor network in 2D, where  the vertical direction is the euclidean time $t$, while the horizontal direction is the spatial direction $x$. b) By using a boundary state we can approximate the full 3D network as the computation of the norm of a 2D iPEPS state. We contract one layer of the tensor network, the transfer matrix $T$, with the boundary iPEPS state $\ket{\psi_0}$ (first line). The bond dimension of the iPEPS state would increase exponentially with the number of step, and thus we approximate the iPEPS state by  truncating its bond dimension back to $D$  (second and third line). We iterate these last two steps many times until the boundary state has actually converged. The converged iPEPS represents our best approximation of the ground state   $\ket{\Omega}$ with finite bond dimension $D$. 
\label{fig:2dnetwork}}
\end{figure}

\begin{figure}[h]
\includegraphics[width=1\columnwidth]{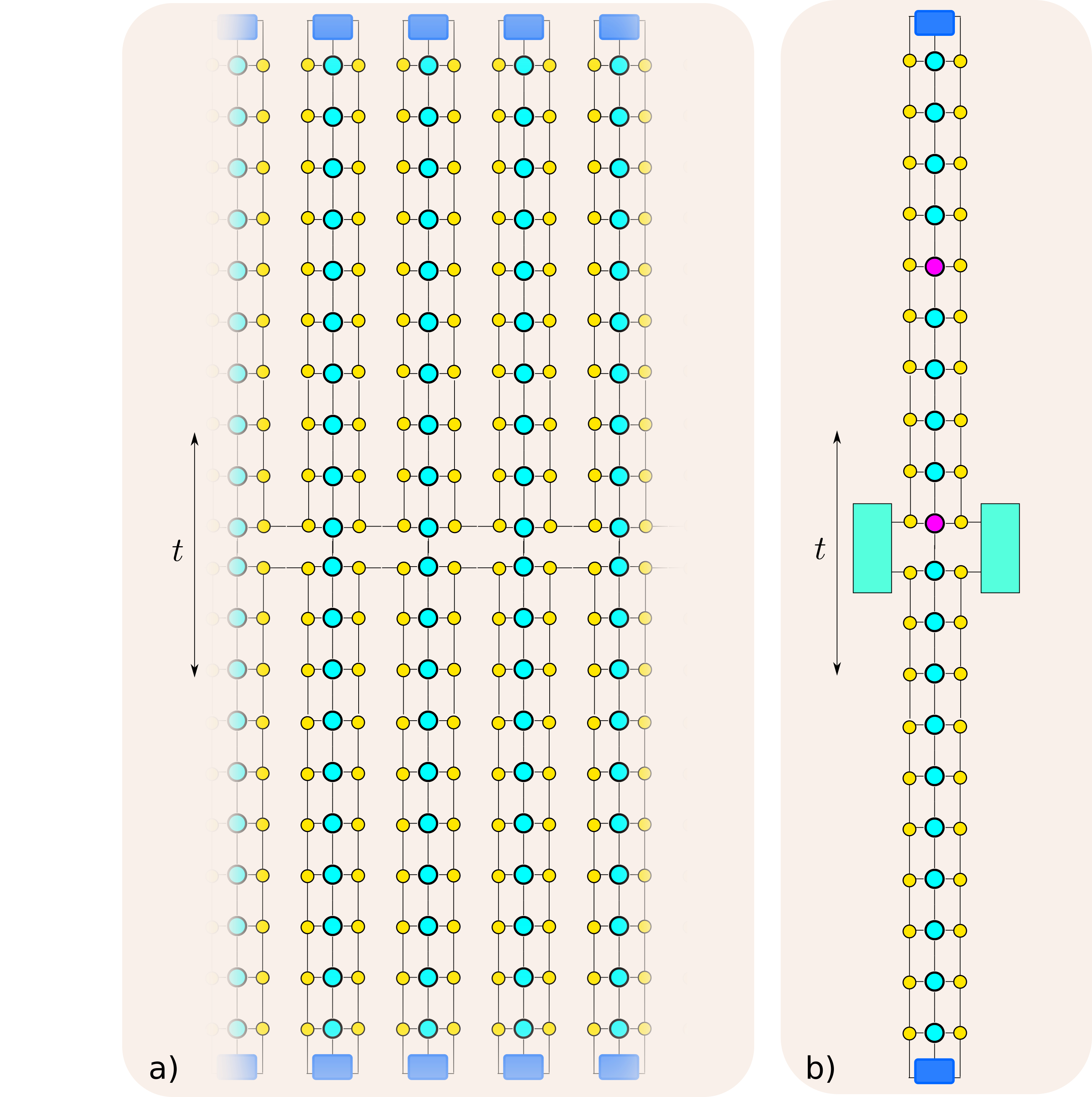}
\caption{(Color online) (a) The fixed point iPEPS tensors are the result of many contraction and projection steps. Here we expand explicitly the iPEPS into its elementary constituents along the imaginary time direction (vertical in the figure), made by the original constituents of the 3D network (cyan tensors) and isometries used at each step to project the evolved iPEPS to a state with bond dimension $D$ (yellow tensors). The result is that the infinite 3D tensor network becomes very anisotropic, and we observe the appearance of quasi one-dimensional time channels. b) The  equal-space, non-equal time correlation function $\braket{\Omega|\mathcal{O}(t_0)\mathcal{O}(t_1)|\Omega}$. The insertion of the local operators $\mathcal{O}$ at two different times $t_0$ and $t_1$ is indicated by purple tensors. The contraction of the network outside the selected time channel produces the two environment tensors represented as rectangles  on the left and the right of the time channel. The correlation function thus assumes a  one dimensional structure. The isometries provide a time-like MPS state. Since the bond dimension of the MPS state is finite and equal to $D$, we expect that these correlations either do not decay or decay exponentially with the time separation $t_1-t_0$.   
\label{fig:2dchannel}}
\end{figure}

The easiest way to understand why a finite $D$ induces a finite correlation time is to represent a 2D cut of the 3D tensor network in which we only represent $x,t$ of the $x,y,t$ coordinates. We will describe the process of encoding the 3D network in a 2D iPEPS as if we were actually performing an imaginary time evolution. The 2D cut of the infinite tensor network is represented in Fig. \ref{fig:2dnetwork}. There, as usual, tensors are represented by geometric shape and the line attached to them represent their indices. Lines connecting two tensors encode the contraction of the two tensors with respect to the specific indices represented by the line.
The elementary tensors of the 3D network  are obtained for example using the procedure described in the previous section  and we assume they have horizontal bond dimension $d$. The transfer matrix $T$ here is represented by a single horizontal line of the network. One way to contract the network is to use boundary states $\ket{\psi_0}$ both at $t=-\infty$ and $t=+\infty$ that are already iPEPS. This strategy allows to encode the full network contraction into  a final iPEPS norm calculation. In  Fig.~\ref{fig:2dnetwork} the boundary states are represented in blue. In the 2D cut the iPEPS boundary state looks like a 1D matrix product state with bond dimension $D$. We can now contract the network from  $t=-\infty$ down to $t=0$ and simultaneously from  $t=+\infty$ up to $t=0$. This is done step by step  by contracting each time the transfer matrix $T$ with the boundary iPEPS that thus can be seen as evolving in imaginary time \cite{vidal_classical_2007,vidal_efficient_2004,orus_infinite_2008}. At each step the bond dimension of the iPEPS increases by a factor $d$, and thus it increases exponentially with time. In order to keep the bond dimension of the boundary iPEPS finite, at each step we need to approximate the iPEPS states with another state with fixed bond dimension $D$. This is done by projecting the tensor product Hilbert space on the horizontal bonds having dimension  $Dd$ back to the original $D$. Practically one needs to find optimal isometries that would perform such truncation (yellow tensors in Fig.~\ref{fig:2dchannel}). For the sake of our argument it is enough to assume that these isometries exist without entering the details on how to obtain them. 

The procedure is repeated until the iPEPS state converges to a fixed point which encodes our best approximation of the ground state with an iPEPS with finite bond dimension $D$. By explicitly representing the action of the isometries on the original tensor network (as first suggested in Fig. 1 of \cite{ferris15}) we realize that the resulting tensor network looks very anisotropic. This is represented in Fig. \ref{fig:2dchannel}. On both panels of Fig. \ref{fig:2dchannel} the isometries are represented by yellow tensors. In panel a) we show that by iteratively projecting the iPEPS states during the imaginary time evolution we are actually creating almost one-dimensional channels along the time direction.  If we now want to study the decay of correlations along the time direction, we need to characterize the decay of $\braket{\Omega|\mathcal{O}(t_0)\mathcal{O}(t_1)|\Omega}$. In panel b) we represent the insertion of the two operators at certain times $t_0$ and $t_1$ by coloring in purple the corresponding tensors in the network. We immediately see that once the remaining part of the network is contracted into an effective environment, (cyan rectangle), the computation is equivalent to a one dimensional computation for the channel along the Euclidean time direction. Along this channel, the system is described by an effective matrix product state with finite bond dimension created by the isometries (panel b) of Fig. \ref{fig:2dchannel}.  

This immediately suggests that equal space, non-equal time correlation functions  decay exponentially, since we already know that 1D quantum system described by iMPS with finite bond dimension cannot be critical. The correlations in those states  either do not decay, or decay exponentially. This means that we expect $\braket{\Omega|\mathcal{O}(t_0)\mathcal{O}(t_1)|\Omega} \propto \exp\left(\frac{-|t_1-t_0|}{\xi_t}\right)$. Since these correlations are generated by powers of $T$ that is a function of $H$, the approximation scheme has effectively introduced a gap into the Hamiltonian of the system. The approximation thus   acts like a relevant perturbation to the critical Hamiltonian \cite{cardy_scaling_1996}. 

Thanks to the theorem proven by Hastings \cite{hastings_spectral_2006}, the ground state of a gapped Hamiltonian has exponentially decaying correlation functions, and this we think is the physical origin of the finite correlation length we observe in our simulations. 

This reasoning seems to suggest that  the best iPEPS approximation to a critical ground state, in the scaling limit, where we can think of our system as a continuous system, transforms a smooth 3D solid geometry, into a "swiss-cheese", in which the size of holes (the valleys between different channels) is of the order of the lattice spacing and the correlation length along the channels actually provides the IR-cutoff inducing all the phenomenology observed numerically. A cartoon of this is sketched in Figure. \ref{fig:towers}.

A legitimate question is how to accommodate in this picture the existence of iPEPS with finite bond dimension and polynomially decaying correlations functions, such as those that are ground state of generalized RK Hamiltonians. These states are known to describe a Lifshitz critical point,  with low energy dispersion relation $E(k)\propto k^z$ with $z\ge 2$ \cite{isakov_dynamics_2011}.  This implies that contrary to what happens at Lorentz invariant critical points, at Lifshitz points space and time play a very different role. In particular these systems embed in 3D the 2D criticality of a classical system where the spatial correlations can be critical even if the imaginary time correlations are cutoff by a finite inverse temperature.

\section{Finite correlation length scaling} \label{sec:fes}
Finite correlation length scaling (also known as finite entanglement scaling) has been introduced and applied in the context of infinite MPS~\cite{tagliacozzo08,pollmann2009,pirvu12} for the study of critical properties of 1D quantum systems, as an alternative to standard finite size scaling using finite MPS. The basic idea is that the finite bond dimension $D$ induces a finite correlation length (a finite amount of entanglement), which can be used as a relevant length scale to perform a scaling analysis, in a very similar way as in the usual finite size scaling approach, i.e. by replacing the system size $L$ by the effective correlation length at criticality $\xid$. For example, for an order parameter $m$ the ansatz reads
\begin{equation}
\label{eq:mansatz}
m(g,L) = L^{-\beta/\nu} {\cal F}(g L^{1/\nu}) \rightarrow \xi_D^{-\beta/\nu} {\cal M}(g \xi_D^{1/\nu}) = m(g,D),
\end{equation}
where $g$ denotes the distance to the critical point and $\xi_D:= \xi(g=0,D)$.

A similar idea has also been used for 2D classical partition functions represented as a 2D tensor network~\cite{nishino96}, which are contracted using the CTM approach, where the  finite boundary dimension $\chi$ introduces an effective correlation length $\xi_\chi$. 

To what extent FCLS can also be applied to 2D quantum systems  using iPEPS has not been explored yet. This generalization is also more challenging because the effective correlation length is affected by both the bond dimension $D$ of the ansatz and the boundary dimension $\chi$ used in the CTM contraction, i.e. in general scaling ansaetze depend on both $\xi_D$ and $\xi_\chi$, e.g. 
\begin{equation}
\label{eq:mansatz2}
m(g,D,\chi) = \xi_D^{-\beta/\nu} {\cal M}(g \xi_D^{1/\nu},\xi_D/\xi_\chi).
\end{equation}
In order to solve this issue we eliminate the $\chi$ dependence by extrapolating the data in $\chi$, and perform a scaling analysis based on $\xi_D := \xi(g=0,D,\chi\rightarrow \infty)$ only, so that the scaling ansatz reduces to Eq.~\ref{eq:mansatz}. In the following we numerically demonstrate  that this scaling ansatz can be used to extract critical properties in the 2D quantum case.

\section{Spinless fermions on the honeycomb lattice} \label{sec:results}
We consider a model of interacting spinless fermions on the honeycomb lattice at half filling, given by the Hamiltonian,
\begin{equation}
 \hat{H} = -t \sum_{\langle \mathbf{i,j} \rangle} \left[ \hat{c}_{\mathbf{i}}^{\dagger} \hat{c}_{\mathbf{j}} + h.c. \right]+ V \sum_{\langle \mathbf{i,j} \rangle} \hat{n}_{\mathbf{i}} \hat{n}_{\mathbf{j}},   %  \label{eq:Ham}
\end{equation}
where the first term describes a nearest-neighbor hopping with amplitude $t$ and the second term a repulsive nearest-neighbor interaction with strength $V$, with $\hat{n_i} = \hat{c}_{\mathbf{i}}^{\dagger} \hat{c}_{\mathbf{i}}$. 
This model has been intensely studied in the past~\cite{wang14,scherer15,li15,wang15b,wang16b,capponi17} and is known to undergo a continuous phase transition between a Dirac semimetal phase and a charge-density wave (CDW) phase at a critical coupling of $V_c/t=1.356(1)$~\cite{wang14}. In the continuum limit the transition belongs to the chiral Ising Gross-Neveu universality class with a dynamical critical exponent $z=1$.

\begin{figure}[h]
\includegraphics[width=1\columnwidth]{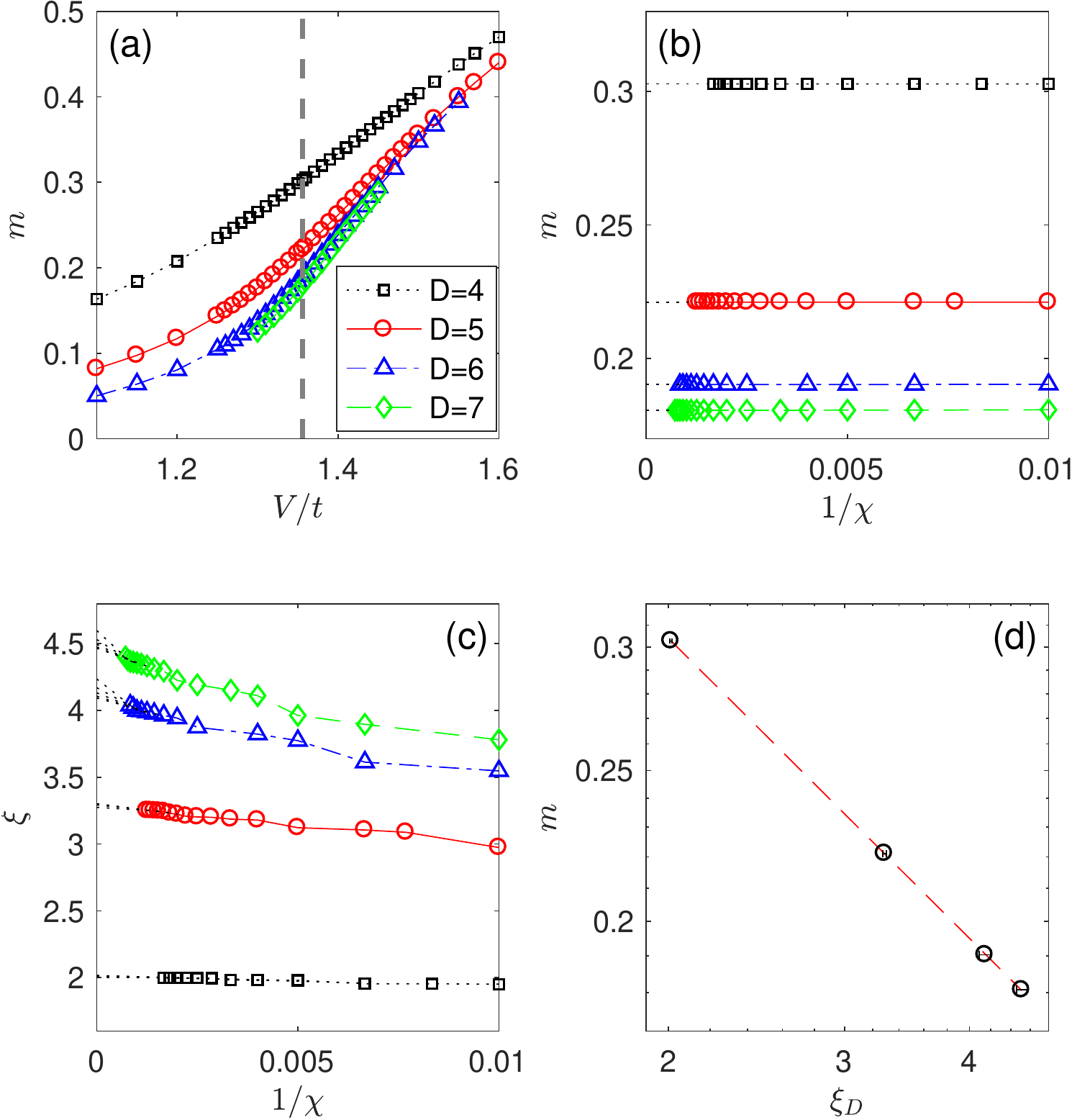}
\caption{(Color online) (a) CDW order parameter as a function of interaction strength $V/t$ for different values of $D$. The location of the critical point is indicated by the dashed line. (b) Order parameter $m$ and (c) correlation length $\xi$  as a function of inverse boundary dimension $\chi$ for different values of $D$ at the critical point $V/t=1.356$. (d) A power-law fit at the critical point yields an exponent $\beta/\nu =0.64(2)$, in agreement with the QMC result.}%which is in agreement with the QMC result $\beta/\nu = 0.651$. } %error between 0.01 and 0.02
\label{fig:crit}
\end{figure}

The CDW order parameter is given by 
\begin{equation}
m = |n_A - n_B|% \in [0,1]
\end{equation}
where $n_A$ and $n_B$ are the particle densities on sublattices $A$ and $B$, respectively. Figure~\ref{fig:crit}(a) shows the order parameter as a function of interaction strength $V/t$ for different bond dimensions $D$, where one can clearly observe the finite $D$ effects, i.e. no sharp phase transition at $V_c/t=1.356$, but  a systematic suppression of $m$ with increasing $D$, very similar to standard finite size effects.

We first test the scaling ansatz at the critical point, i.e. $g=(V-V_c)/V_c=0$, with $V_c/t=1.356$, where Eq.~\ref{eq:mansatz2} reduces to
\begin{equation}
\label{eq:mansatz}
m(g=0,D,\chi) =  \xid^{-\beta/\nu} {\cal M}(0 \cdot \xi_D^{1/\nu},\xic/\xid)   \sim  \xid^{-\beta/\nu}.
\end{equation}
The latter relation is obtained by taking a sufficiently large $\chi$ such that $m$ is fully converged in $\chi$. Fig.~\ref{fig:crit}(b) shows that $m$ converges rapidly in $\chi$ such that no extrapolation in $\chi$ is needed. However, the correlation length $\xi$ displays a stronger dependence on $\chi$~\footnote{The correlation length is computed from the largest and second largest eigenvalues of the row-to-row transfer matrix as in Ref.~\cite{nishino96}}, shown in Fig.~\ref{fig:crit}(c). We determine $\xid$ for each value of $D$ by extrapolating $\xi(D,\chi)$ to the infinite $\chi$ limit. We do this by performing linear extrapolations in $1/\chi$ using different ranges of data points, and determining the average and standard deviation of these extrapolations~\footnote{During completion of this work an interesting alternative approach to obtain the correlation length in the infinite~$\chi$ limit was introduced in Ref.~\cite{rams18}}. We clearly find that even at the largest bond dimension the state is not critical, i.e. that the finite $D$ induces a finite correlation length, similarly as in 1D with MPS. 
Finally,  Fig.~\ref{fig:crit}(d) shows a log-log plot of $m$ versus $\xid$, where a linear fit yields an estimate for $\beta/\nu =0.64(2)$  in agreement with the QMC result $\beta/\nu = 0.65(4)$ from Ref.~\cite{wang14}.

%%%%%%%%%%%%%%%%%%%%%%%%%%%
%%%%%%%%%%%%%%%%%%%%%%%%%%%

\begin{figure}
\includegraphics[width=1\columnwidth]{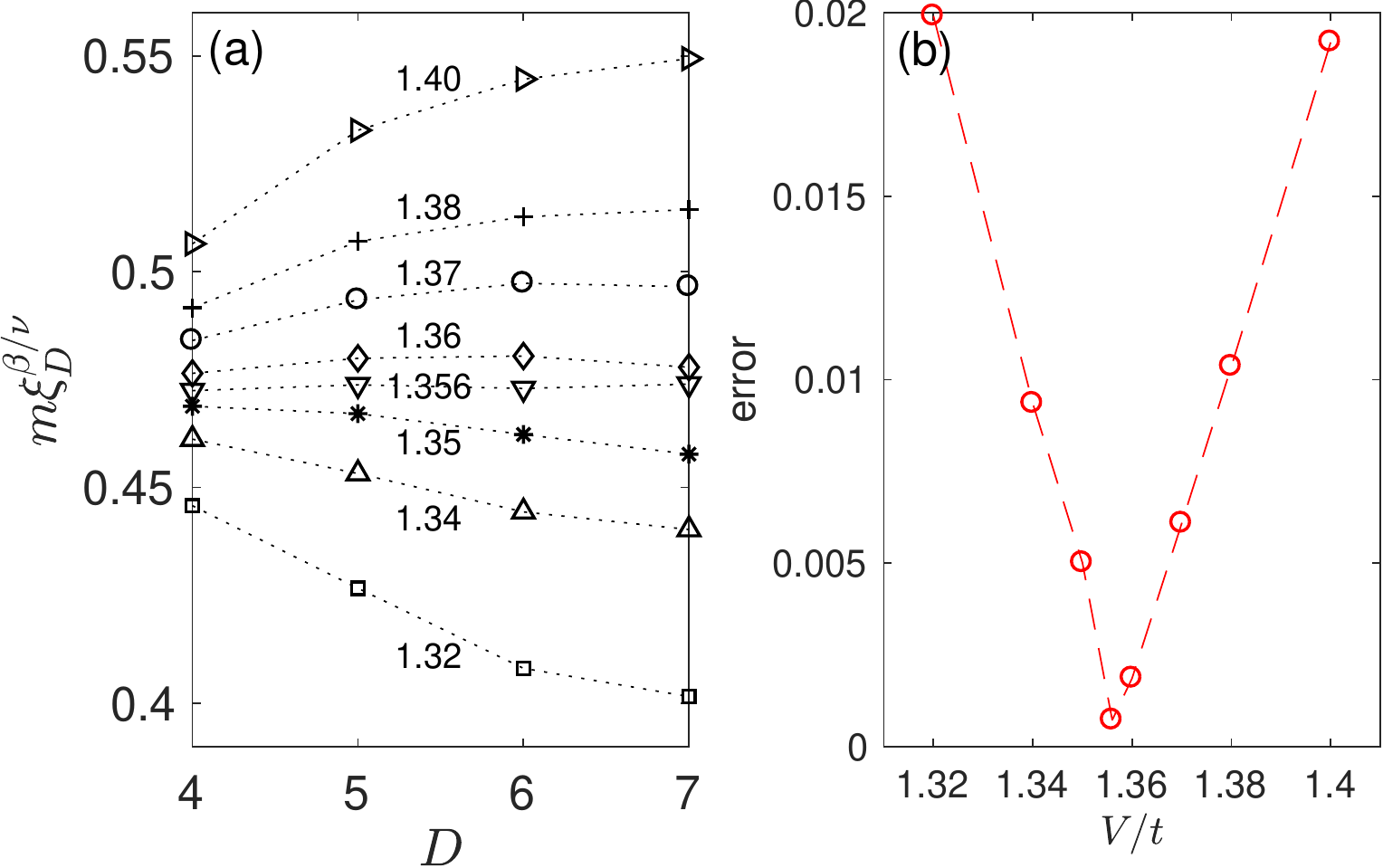}
\caption{(Color online) Results for $y=m \xi^{\beta/\nu}$ as a function of $D$ for fixed $\beta/\nu=0.64$. At the critical point $V_c/t=1.356$ the value of $y$ becomes independent of the bond dimension $D$, cf. Eq.~\ref{eq:fixedbeta}. The standard deviation of the points is shown on the right hand side, which clearly exhibits a minimum at the critical point.
}
\label{fig:findcrit}
\end{figure}

Next, we check if we can consistently determine the critical point $V_c$  based on this result for $\beta/\nu$. Using the scaling ansatz \eqref{eq:mansatz2} at the critical point in the large $\chi$ limit we obtain 
\begin{equation}
\label{eq:fixedbeta}
y=m(g=0,D,\chi\rightarrow \infty) \, \xid^{\beta/\nu} = const,
\end{equation}
i.e. a constant independent of $D$. In Fig.~\ref{fig:findcrit}(a) we plot $y$ as a function of $D$ for different values of $V/t$ and Fig.~\ref{fig:findcrit}(b) shows the deviation from a straight line (computed by the standard deviation of $y$) as a function of $V/t$ (where $\xi_D$ is determined for each value of $V/t$). The smallest deviation is obtained for $V/t=1.356$, consistent with the QMC result. (A similar result is also obtained using $\beta/\nu=0.65$).

%%%%%%%%%%%%%%%%%%%%%%%%%%%
%%%%%%%%%%%%%%%%%%%%%%%%%%%

\begin{figure}
\includegraphics[width=1\columnwidth]{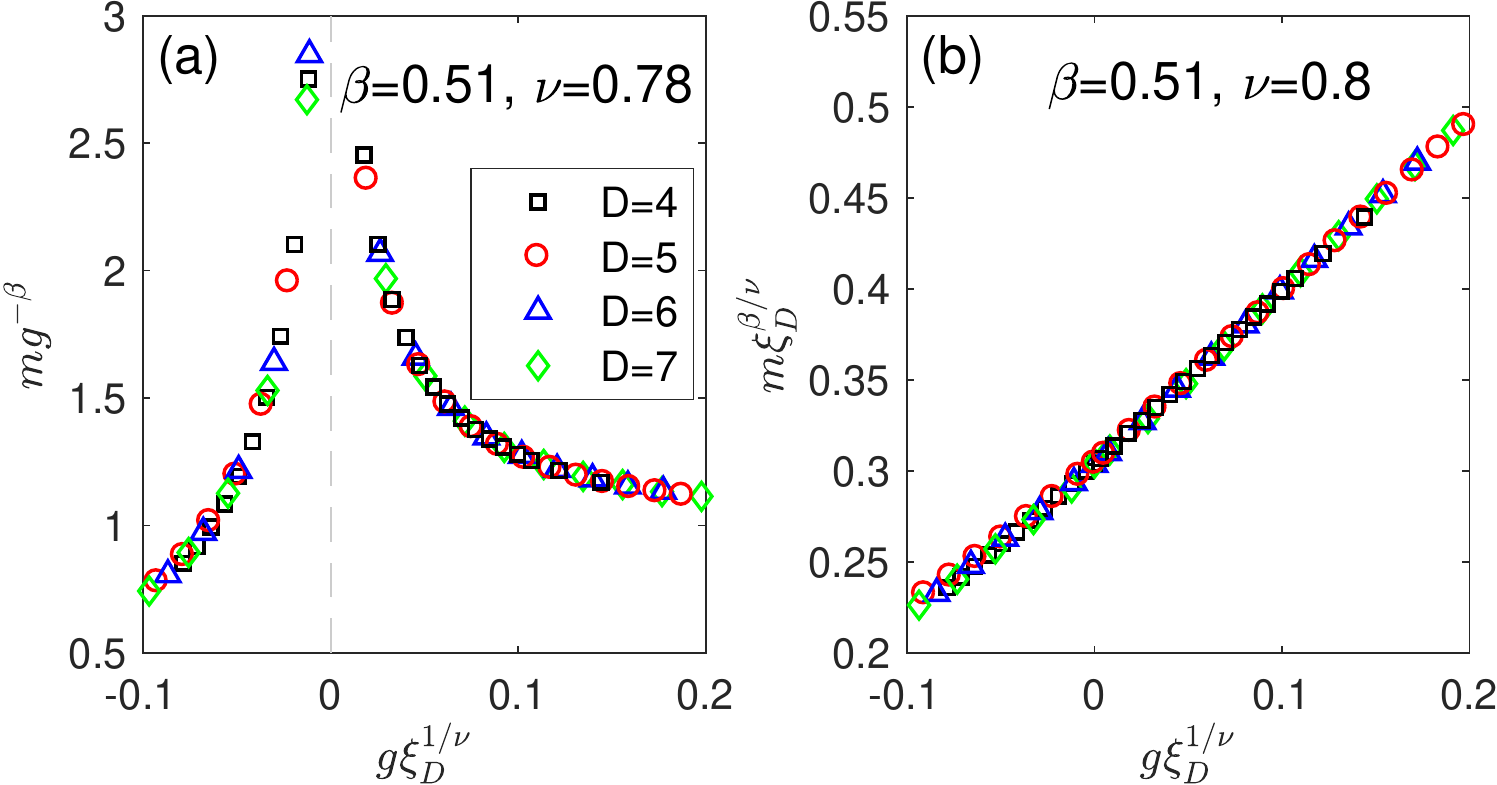}
\caption{(Color online) Data collapse plots using Eqs.~\ref{eq:collapse} and \ref{eq:mgansatz}.
}
\label{fig:collapse}
\end{figure}

We next attempt to perform a data collapse using the following two ansaetze (again in the large $\chi$ limit),
\begin{eqnarray}
\label{eq:collapse}
%y_1 &=& 
m(g,D) \, \xid^{\beta/\nu} &=& {\cal M}(g\xid^{1/\nu}),\\
%y_2 &=& 
m(g,D) \, g^{-\beta} &=& {\cal \tilde M}(g\xid^{1/\nu}),
\label{eq:mgansatz}
\end{eqnarray}
to determine the critical exponents $\nu$ and $\beta$ (with $V_c/t=1.356$), shown in Fig.~\ref{fig:collapse}. By varying the range of data points and by taking the error bar of $\xid$ into account, we obtain the following estimates: $\beta=0.51(1)$, and $\nu=0.79(2)$ in agreement with the QMC result $\beta=0.52(3)$, $\nu=0.80(3)$~\cite{wang14}.

%%%%%%%%%%%%%%%%%%%%%%%%%%%
%%%%%%%%%%%%%%%%%%%%%%%%%%%

%\emph{Determining the critical point --}
\subsection{Determining the critical point based on $m'/m$}
\begin{figure}
\includegraphics[width=1\columnwidth]{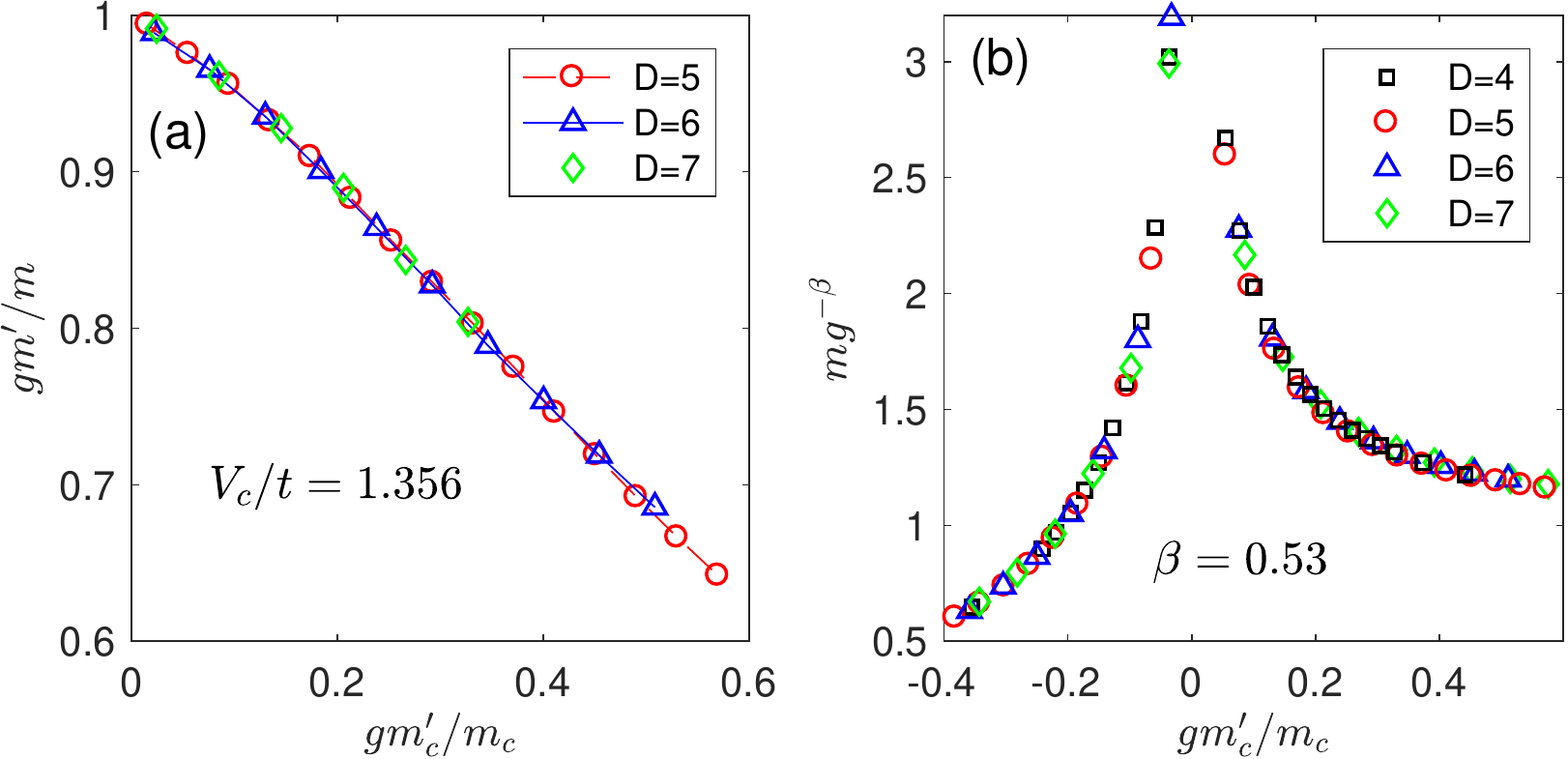}
\caption{(a) Data collapse using Eq.~\eqref{eq:collapseVc} yielding $V_c/t=1.356(2)$. (b) Data collapse based on Eq.~\eqref{eq:mgansatz2} giving $\beta=0.53(1)$.}
\label{fig:collapseVc}
\end{figure}

A standard method to determine the critical point without knowledge of the critical exponents is by the Binder cumulant, which is invariant at the critical point for different system sizes (or bond dimensions). However, this would require computing the fourth order moment of the order parameter, which is difficult in a 2D tensor network approach. Here we introduce an alternative approach based on the derivative of the order parameter with respect to $g$, which in the large $\chi$ limit is expected to obey the following ansatz

\begin{equation}
\label{eq:xi2}
m'(g,D) =  \xid^{-(\beta-1)/\nu} {\cal M'}(g \xi_D^{1/\nu}).  
\end{equation}
Thus at the critical point, $g=0$,
\begin{equation}
\frac{m'_c(D)}{m_c(D)}  := \frac{m'(g=0,D)}{m(g=0,D)} \sim \xi_D^{1/\nu}.
\end{equation}
Thus we have found an expression for $\xi_D^{1/\nu}$ which we can use to rewrite scaling functions:
\begin{equation}
\label{eq:x}
{\cal M}(g \xi_D^{1/\nu}) \sim {\cal P}\left(g\frac{m'_c(D)}{m_c(D)} \right).
\end{equation}
Dividing Eq.~\eqref{eq:xi2} by $m$ and multiplying by $g$ yields 
\begin{equation}
g \frac{m'(g,D)}{m(g,D)} = {\cal  \bar M'}(g \xi_D^{1/\nu}) =  {\cal P}\left(g\frac{m'_c(D)}{m_c(D)}  \right).\\
%g\frac{m'(g,D,\chi)}{m(g,D,\chi)} &=& g\xi_D^{1/\nu} {\cal M'''}(g \xi_D^{1/\nu}) = {\cal M''''}(g \xi_D^{1/\nu}) 
\label{eq:collapseVc}
\end{equation}
Thus, we can plot $y=(V-V_c) \frac{m'(g,D)}{m(g,D)} $ versus $x=(V-V_c)\frac{m'_c(D)}{m_c(D)}$ as a function of $V$, and for the correct choice of $V_c$ the data for different values of $D$  collapse onto a single curve. The best data collapse, shown in Fig.~\ref{fig:collapseVc}(a), is obtained for $V_c/t=1.356(2)$ in agreement with the QMC result. The numerical derivative has been obtained by taking the derivative of polynomial fits to $m$ versus $V/t$ data. 

%%%%%%%%%%%%%%%%%%%%%%%%%%%
%%%%%%%%%%%%%%%%%%%%%%%%%%%

Finally, we can combine Eq.~\ref{eq:x} and Eq.~\ref{eq:mgansatz},
\begin{equation}
\label{eq:mgansatz2}
m(g,D) \, g^{-\beta} = {\cal \tilde P}\left(g\frac{m'_c(D)}{m_c(D)} \right),
\end{equation}
 to perform a data collapse and extract the exponent~$\beta$, shown in Fig.~\ref{fig:collapseVc}, yielding a consistent value of $\beta=0.53(1)$.

\section{Improved extrapolations of order parameters}
\label{sec:op}
\begin{figure}
\includegraphics[width=1\columnwidth]{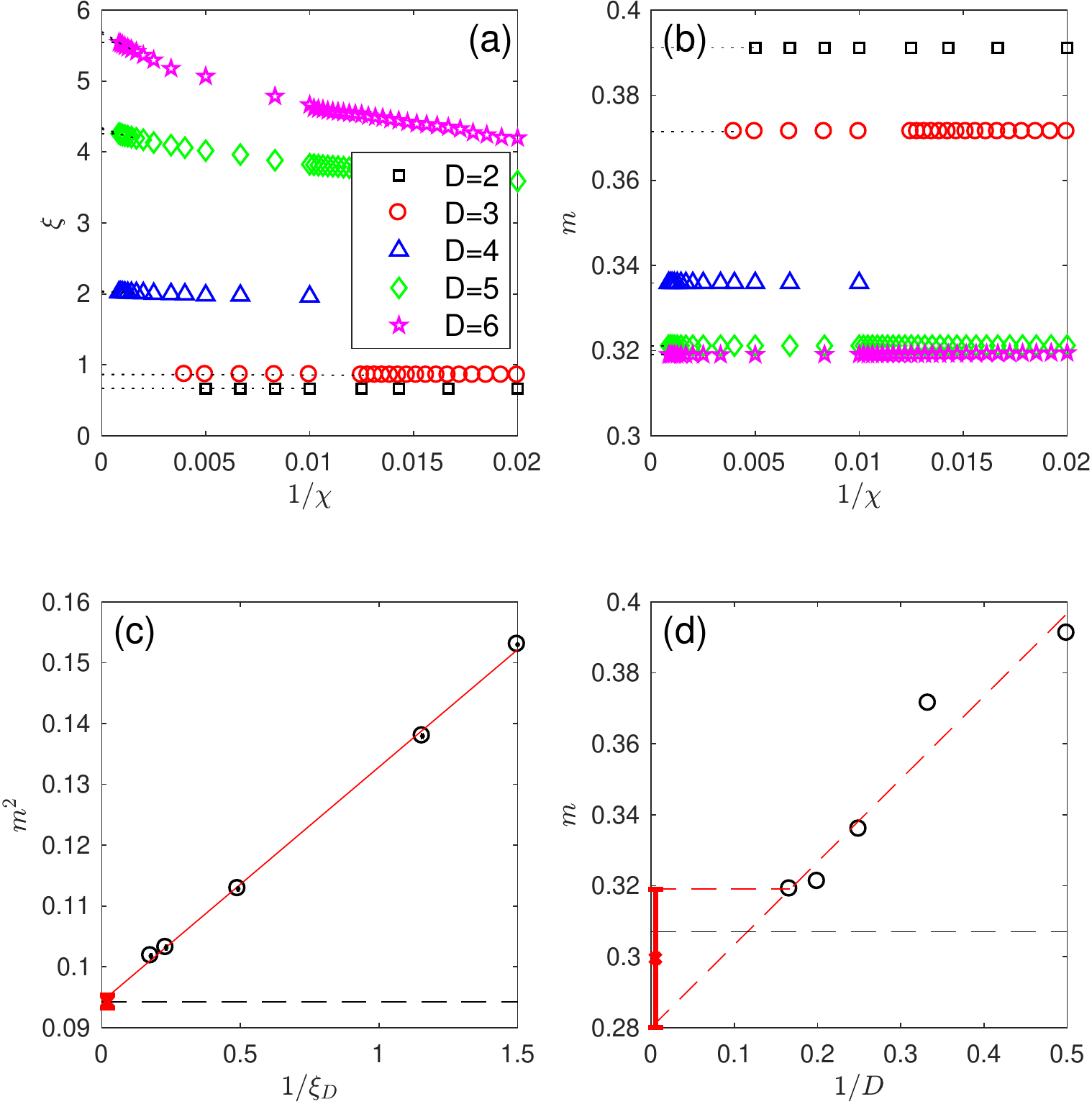}
\caption{(Color online) iPEPS results for the 2D Heisenberg model. (a) Correlation length $\xi$ and (b) staggered magnetization $m$ as a function of inverse  $\chi$ for different values of $D$. (c) Squared of the staggered magnetization as a function of inverse $\xi_D$. A linear extrapolation yields $m=0.307 \pm 0.002$ which is in agreement with the QMC result $m = 0.30743(1)$. (d) Order parameter as a function of inverse $D$.}
\label{fig:heis}
\end{figure}

The effective correlation length $\xid$ can also be used to perform an improved extrapolation of the order parameter in a gapless system. As an illustration we present results for the two-dimensional $S=1/2$ Heisenberg model on a square lattice.  Fig.~\ref{fig:heis}(a) and (b) show the correlation length $\xi$ and staggered magnetization $m$ as a function of inverse $\chi$, respectively. 
A linear extrapolation of $m^2$ as a function of $\xid$ in Fig.~\ref{fig:heis}(c)  yields $m=0.307 \pm 0.002$, in agreement with the state-of-the-art QMC result $m = 0.30743(1)$ from Ref.~\cite{Sandvik10}. This approach provides a much better estimate than the one based on a crude $1/D$ extrapolation shown in Fig.~\ref{fig:heis}(d), which is not very accurate due to the non-monotonic behavior as a function of $D$.

%%%%%%%%%%%%%%%%%%%%%%%%%%%
%%%%%%%%%%%%%%%%%%%%%%%%%%%

\section{Summary and conclusion} \label{sec:conclusion}
In this paper we have demonstrated the usefulness and applicability of finite correlation length scaling in two dimensions based on iPEPS by determining the critical exponents and critical point of an interacting spinless fermion model on a honeycomb lattice. Our findings are in agreement with the QMC results from Ref.~\cite{wang14}. 
Furthermore, we introduced a new approach to determine the critical point based on the derivative of the order parameter, which does not  require the computation higher order moments of the order parameter or extrapolations of the effective correlation length in $\chi$, making it a particular useful approach for 2D tensor network calculations.

We stress that  iPEPS  can also be applied to models which are inaccessible to QMC due to the negative sign problem, making it a promising tool to  study critical properties of challenging open problems.

From the theoretical point of view, we have possibly identified a class of models whose ground states are hard to encode with iPEPS with finite bond dimension, despite obeying the area-law of  entanglement. We have given an intuitive argument that the mismatch between the exact critical states and the finitely correlated iPEPS  we obtain has a geometric origin. The iPEPS seems to approximate the smooth continuous 3D geometry with a landscape of towers separated by valleys, where the finite correlation time along the towers could be the ultimate reason for the appearance of the finite correlation length.  

Our results are also relevant to field theories. The operator content of a field theory is known to depend on the geometry and the boundary conditions of the space on which the theory is defined \cite{cardy_operator_1986, cardy_effect_1986}. It would be important to understand the effects of the landscape of towers and valleys on the operator content of the corresponding field theory.
Furthermore, our results seem to suggest that $D$ would act as a regulator in any continuum limit of the theory defined on the lattice.  The bond dimension of the iPEPS should indeed increase in order to keep the ratio between the relevant physical quantities and the correlation length fixed as we approach the continuum limit.

{\em Note:} Similar results have been reported by M.~Rader and A.~L\"auchli, see arXiv:1803.?????.

\begin{acknowledgments}
We acknowledge  discussions with Andreas L\"auchli and Michael Rader. LT acknowledges the long-time discussions and collaboration with Andrea Coser on similar problems studied in the context of Gaussian tensor networks. This project was supported by the European Research Council (ERC) under the European UnionÕs Horizon 2020 research and innovation programme (grant agreement No 677061), and by the Narodowe Centrum Nauki (National Science Center) under Project No. 2016/23/B/ST3/00830. This work is part of the \mbox{Delta-ITP} consortium, a program of the Netherlands Organization for Scientific Research (NWO) that is funded by the Dutch Ministry of Education, Culture and Science~(OCW).
\end{acknowledgments}

\bibliographystyle{apsrev4-1}
\bibliography{refs,fes_peps}

%merlin.mbs apsrev4-1.bst 2010-07-25 4.21a (PWD, AO, DPC) hacked
%Control: key (0)
%Control: author (72) initials jnrlst
%Control: editor formatted (1) identically to author
%Control: production of article title (-1) disabled
%Control: page (0) single
%Control: year (1) truncated
%Control: production of eprint (0) enabled
\begin{thebibliography}{80}%
\makeatletter
\providecommand \@ifxundefined [1]{%
 \@ifx{#1\undefined}
}%
\providecommand \@ifnum [1]{%
 \ifnum #1\expandafter \@firstoftwo
 \else \expandafter \@secondoftwo
 \fi
}%
\providecommand \@ifx [1]{%
 \ifx #1\expandafter \@firstoftwo
 \else \expandafter \@secondoftwo
 \fi
}%
\providecommand \natexlab [1]{#1}%
\providecommand \enquote  [1]{``#1''}%
\providecommand \bibnamefont  [1]{#1}%
\providecommand \bibfnamefont [1]{#1}%
\providecommand \citenamefont [1]{#1}%
\providecommand \href@noop [0]{\@secondoftwo}%
\providecommand \href [0]{\begingroup \@sanitize@url \@href}%
\providecommand \@href[1]{\@@startlink{#1}\@@href}%
\providecommand \@@href[1]{\endgroup#1\@@endlink}%
\providecommand \@sanitize@url [0]{\catcode `\\12\catcode `\$12\catcode
  `\&12\catcode `\#12\catcode `\^12\catcode `\_12\catcode `\%12\relax}%
\providecommand \@@startlink[1]{}%
\providecommand \@@endlink[0]{}%
\providecommand \url  [0]{\begingroup\@sanitize@url \@url }%
\providecommand \@url [1]{\endgroup\@href {#1}{\urlprefix }}%
\providecommand \urlprefix  [0]{URL }%
\providecommand \Eprint [0]{\href }%
\providecommand \doibase [0]{http://dx.doi.org/}%
\providecommand \selectlanguage [0]{\@gobble}%
\providecommand \bibinfo  [0]{\@secondoftwo}%
\providecommand \bibfield  [0]{\@secondoftwo}%
\providecommand \translation [1]{[#1]}%
\providecommand \BibitemOpen [0]{}%
\providecommand \bibitemStop [0]{}%
\providecommand \bibitemNoStop [0]{.\EOS\space}%
\providecommand \EOS [0]{\spacefactor3000\relax}%
\providecommand \BibitemShut  [1]{\csname bibitem#1\endcsname}%
\let\auto@bib@innerbib\@empty
%</preamble>
\bibitem [{\citenamefont {Verstraete}\ \emph {et~al.}(2008)\citenamefont
  {Verstraete}, \citenamefont {Murg},\ and\ \citenamefont
  {Cirac}}]{Verstraete08}%
  \BibitemOpen
  \bibfield  {author} {\bibinfo {author} {\bibfnamefont {F.}~\bibnamefont
  {Verstraete}}, \bibinfo {author} {\bibfnamefont {V.}~\bibnamefont {Murg}}, \
  and\ \bibinfo {author} {\bibfnamefont {J.~I.}\ \bibnamefont {Cirac}},\ }\href
  {\doibase 10.1080/14789940801912366} {\bibfield  {journal} {\bibinfo
  {journal} {Advances in Physics}\ }\textbf {\bibinfo {volume} {57}},\ \bibinfo
  {pages} {143} (\bibinfo {year} {2008})}\BibitemShut {NoStop}%
\bibitem [{\citenamefont {Schollw\"ock}(2011)}]{schollwoeck2011}%
  \BibitemOpen
  \bibfield  {author} {\bibinfo {author} {\bibfnamefont {U.}~\bibnamefont
  {Schollw\"ock}},\ }\href {\doibase 10.1016/j.aop.2010.09.012} {\bibfield
  {journal} {\bibinfo  {journal} {Annals of Physics}\ }\textbf {\bibinfo
  {volume} {326}},\ \bibinfo {pages} {96} (\bibinfo {year} {2011})}\BibitemShut
  {NoStop}%
\bibitem [{\citenamefont {Or{\'u}s}(2014)}]{orus_practical_2014}%
  \BibitemOpen
  \bibfield  {author} {\bibinfo {author} {\bibfnamefont {R.}~\bibnamefont
  {Or{\'u}s}},\ }\href {\doibase 10.1016/j.aop.2014.06.013} {\bibfield
  {journal} {\bibinfo  {journal} {Annals of Physics}\ }\textbf {\bibinfo
  {volume} {349}},\ \bibinfo {pages} {117} (\bibinfo {year}
  {2014})}\BibitemShut {NoStop}%
\bibitem [{\citenamefont {Bridgeman}\ and\ \citenamefont
  {Chubb}(2017)}]{bridgeman_hand-waving_2017}%
  \BibitemOpen
  \bibfield  {author} {\bibinfo {author} {\bibfnamefont {J.~C.}\ \bibnamefont
  {Bridgeman}}\ and\ \bibinfo {author} {\bibfnamefont {C.~T.}\ \bibnamefont
  {Chubb}},\ }\href {\doibase 10.1088/1751-8121/aa6dc3} {\bibfield  {journal}
  {\bibinfo  {journal} {J. Phys. A: Math. Theor.}\ }\textbf {\bibinfo {volume}
  {50}},\ \bibinfo {pages} {223001} (\bibinfo {year} {2017})}\BibitemShut
  {NoStop}%
\bibitem [{\citenamefont {Haegeman}\ and\ \citenamefont
  {Verstraete}(2017)}]{haegeman_diagonalizing_2017}%
  \BibitemOpen
  \bibfield  {author} {\bibinfo {author} {\bibfnamefont {J.}~\bibnamefont
  {Haegeman}}\ and\ \bibinfo {author} {\bibfnamefont {F.}~\bibnamefont
  {Verstraete}},\ }\href {\doibase 10.1146/annurev-conmatphys-031016-025507}
  {\bibfield  {journal} {\bibinfo  {journal} {Annu. Rev. Condens. Matter
  Phys.}\ }\textbf {\bibinfo {volume} {8}},\ \bibinfo {pages} {355} (\bibinfo
  {year} {2017})}\BibitemShut {NoStop}%
\bibitem [{\citenamefont {Hastings}(2007)}]{hastings_area_2007}%
  \BibitemOpen
  \bibfield  {author} {\bibinfo {author} {\bibfnamefont {M.~B.}\ \bibnamefont
  {Hastings}},\ }\href {\doibase 10.1088/1742-5468/2007/08/P08024} {\bibfield
  {journal} {\bibinfo  {journal} {Journal of Statistical Mechanics: Theory and
  Experiment}\ }\textbf {\bibinfo {volume} {2007}},\ \bibinfo {pages} {P08024}
  (\bibinfo {year} {2007})}\BibitemShut {NoStop}%
\bibitem [{\citenamefont {Wolf}\ \emph {et~al.}(2008)\citenamefont {Wolf},
  \citenamefont {Verstraete}, \citenamefont {Hastings},\ and\ \citenamefont
  {Cirac}}]{wolf_area_2008}%
  \BibitemOpen
  \bibfield  {author} {\bibinfo {author} {\bibfnamefont {M.~M.}\ \bibnamefont
  {Wolf}}, \bibinfo {author} {\bibfnamefont {F.}~\bibnamefont {Verstraete}},
  \bibinfo {author} {\bibfnamefont {M.~B.}\ \bibnamefont {Hastings}}, \ and\
  \bibinfo {author} {\bibfnamefont {J.~I.}\ \bibnamefont {Cirac}},\ }\href
  {\doibase 10.1103/PhysRevLett.100.070502} {\bibfield  {journal} {\bibinfo
  {journal} {Phys. Rev. Lett.}\ }\textbf {\bibinfo {volume} {100}},\ \bibinfo
  {pages} {070502} (\bibinfo {year} {2008})}\BibitemShut {NoStop}%
\bibitem [{\citenamefont {Eisert}\ \emph {et~al.}(2008)\citenamefont {Eisert},
  \citenamefont {Cramer},\ and\ \citenamefont {Plenio}}]{eisert_area_2008}%
  \BibitemOpen
  \bibfield  {author} {\bibinfo {author} {\bibfnamefont {J.}~\bibnamefont
  {Eisert}}, \bibinfo {author} {\bibfnamefont {M.}~\bibnamefont {Cramer}}, \
  and\ \bibinfo {author} {\bibfnamefont {M.~B.}\ \bibnamefont {Plenio}},\
  }\href {http://arxiv.org/abs/0808.3773} {\bibfield  {journal} {\bibinfo
  {journal} {0808.3773}\ } (\bibinfo {year} {2008})}\BibitemShut {NoStop}%
\bibitem [{\citenamefont {Amico}\ \emph {et~al.}(2008)\citenamefont {Amico},
  \citenamefont {Fazio}, \citenamefont {Osterloh},\ and\ \citenamefont
  {Vedral}}]{amico_entanglement_2008}%
  \BibitemOpen
  \bibfield  {author} {\bibinfo {author} {\bibfnamefont {L.}~\bibnamefont
  {Amico}}, \bibinfo {author} {\bibfnamefont {R.}~\bibnamefont {Fazio}},
  \bibinfo {author} {\bibfnamefont {A.}~\bibnamefont {Osterloh}}, \ and\
  \bibinfo {author} {\bibfnamefont {V.}~\bibnamefont {Vedral}},\ }\href
  {\doibase 10.1103/RevModPhys.80.517} {\bibfield  {journal} {\bibinfo
  {journal} {Rev. Mod. Phys.}\ }\textbf {\bibinfo {volume} {80}},\ \bibinfo
  {pages} {517} (\bibinfo {year} {2008})}\BibitemShut {NoStop}%
\bibitem [{\citenamefont {Masanes}(2009)}]{masanes_area_2009}%
  \BibitemOpen
  \bibfield  {author} {\bibinfo {author} {\bibfnamefont {L.}~\bibnamefont
  {Masanes}},\ }\href {\doibase 10.1103/PhysRevA.80.052104} {\bibfield
  {journal} {\bibinfo  {journal} {Phys. Rev. A}\ }\textbf {\bibinfo {volume}
  {80}},\ \bibinfo {pages} {052104} (\bibinfo {year} {2009})}\BibitemShut
  {NoStop}%
\bibitem [{\citenamefont {Laflorencie}(2016)}]{laflorencie_quantum_2016}%
  \BibitemOpen
  \bibfield  {author} {\bibinfo {author} {\bibfnamefont {N.}~\bibnamefont
  {Laflorencie}},\ }\href {\doibase 10.1016/j.physrep.2016.06.008} {\bibfield
  {journal} {\bibinfo  {journal} {Physics Reports}\ }\bibinfo {series} {Quantum
  entanglement in condensed matter systems},\ \textbf {\bibinfo {volume}
  {646}},\ \bibinfo {pages} {1} (\bibinfo {year} {2016})}\BibitemShut {NoStop}%
\bibitem [{\citenamefont {Affleck}\ \emph {et~al.}(1987)\citenamefont
  {Affleck}, \citenamefont {Kennedy}, \citenamefont {Lieb},\ and\ \citenamefont
  {Tasaki}}]{affleck_rigorous_1987}%
  \BibitemOpen
  \bibfield  {author} {\bibinfo {author} {\bibfnamefont {I.}~\bibnamefont
  {Affleck}}, \bibinfo {author} {\bibfnamefont {T.}~\bibnamefont {Kennedy}},
  \bibinfo {author} {\bibfnamefont {E.~H.}\ \bibnamefont {Lieb}}, \ and\
  \bibinfo {author} {\bibfnamefont {H.}~\bibnamefont {Tasaki}},\ }\href
  {\doibase 10.1103/PhysRevLett.59.799} {\bibfield  {journal} {\bibinfo
  {journal} {Phys. Rev. Lett.}\ }\textbf {\bibinfo {volume} {59}},\ \bibinfo
  {pages} {799} (\bibinfo {year} {1987})},\ \bibinfo {note} {copyright (C) 2009
  The American Physical Society; Please report any problems to
  prola@aps.org}\BibitemShut {NoStop}%
\bibitem [{\citenamefont {Ostlund}\ and\ \citenamefont
  {Rommer}(1995)}]{ostlund_thermodynamic_1995}%
  \BibitemOpen
  \bibfield  {author} {\bibinfo {author} {\bibfnamefont {S.}~\bibnamefont
  {Ostlund}}\ and\ \bibinfo {author} {\bibfnamefont {S.}~\bibnamefont
  {Rommer}},\ }\href {\doibase 10.1103/PhysRevLett.75.3537} {\bibfield
  {journal} {\bibinfo  {journal} {Phys. Rev. Lett.}\ }\textbf {\bibinfo
  {volume} {75}},\ \bibinfo {pages} {3537} (\bibinfo {year} {1995})},\ \bibinfo
  {note} {copyright (C) 2009 The American Physical Society; Please report any
  problems to prola@aps.org}\BibitemShut {NoStop}%
\bibitem [{\citenamefont {Perez-Garcia}\ \emph {et~al.}(2007)\citenamefont
  {Perez-Garcia}, \citenamefont {Verstraete}, \citenamefont {Wolf},\ and\
  \citenamefont {Cirac}}]{perez-garcia_matrix_2007}%
  \BibitemOpen
  \bibfield  {author} {\bibinfo {author} {\bibfnamefont {D.}~\bibnamefont
  {Perez-Garcia}}, \bibinfo {author} {\bibfnamefont {F.}~\bibnamefont
  {Verstraete}}, \bibinfo {author} {\bibfnamefont {M.~M.}\ \bibnamefont
  {Wolf}}, \ and\ \bibinfo {author} {\bibfnamefont {J.~I.}\ \bibnamefont
  {Cirac}},\ }\href {http://dl.acm.org/citation.cfm?id=2011832.2011833}
  {\bibfield  {journal} {\bibinfo  {journal} {Quantum Info. Comput.}\ }\textbf
  {\bibinfo {volume} {7}},\ \bibinfo {pages} {401} (\bibinfo {year}
  {2007})}\BibitemShut {NoStop}%
\bibitem [{\citenamefont {Verstraete}\ and\ \citenamefont
  {Cirac}(2004)}]{verstraete2004}%
  \BibitemOpen
  \bibfield  {author} {\bibinfo {author} {\bibfnamefont {F.}~\bibnamefont
  {Verstraete}}\ and\ \bibinfo {author} {\bibfnamefont {J.~I.}\ \bibnamefont
  {Cirac}},\ }\href {http://arxiv.org/abs/cond-mat/0407066} {\bibfield
  {journal} {\bibinfo  {journal} {arXiv:cond-mat/0407066}\ } (\bibinfo {year}
  {2004})}\BibitemShut {NoStop}%
\bibitem [{\citenamefont {Jordan}\ \emph {et~al.}(2008)\citenamefont {Jordan},
  \citenamefont {Or\'{u}s}, \citenamefont {Vidal}, \citenamefont {Verstraete},\
  and\ \citenamefont {Cirac}}]{jordan2008}%
  \BibitemOpen
  \bibfield  {author} {\bibinfo {author} {\bibfnamefont {J.}~\bibnamefont
  {Jordan}}, \bibinfo {author} {\bibfnamefont {R.}~\bibnamefont {Or\'{u}s}},
  \bibinfo {author} {\bibfnamefont {G.}~\bibnamefont {Vidal}}, \bibinfo
  {author} {\bibfnamefont {F.}~\bibnamefont {Verstraete}}, \ and\ \bibinfo
  {author} {\bibfnamefont {J.~I.}\ \bibnamefont {Cirac}},\ }\href {\doibase
  10.1103/PhysRevLett.101.250602} {\bibfield  {journal} {\bibinfo  {journal}
  {Phys. Rev. Lett.}\ }\textbf {\bibinfo {volume} {101}},\ \bibinfo {pages}
  {250602} (\bibinfo {year} {2008})}\BibitemShut {NoStop}%
\bibitem [{\citenamefont {Nishino}\ \emph {et~al.}(2001)\citenamefont
  {Nishino}, \citenamefont {Hieida}, \citenamefont {Okunishi}, \citenamefont
  {Maeshima}, \citenamefont {Akutsu},\ and\ \citenamefont
  {Gendiar}}]{nishino01}%
  \BibitemOpen
  \bibfield  {author} {\bibinfo {author} {\bibfnamefont {T.}~\bibnamefont
  {Nishino}}, \bibinfo {author} {\bibfnamefont {Y.}~\bibnamefont {Hieida}},
  \bibinfo {author} {\bibfnamefont {K.}~\bibnamefont {Okunishi}}, \bibinfo
  {author} {\bibfnamefont {N.}~\bibnamefont {Maeshima}}, \bibinfo {author}
  {\bibfnamefont {Y.}~\bibnamefont {Akutsu}}, \ and\ \bibinfo {author}
  {\bibfnamefont {A.}~\bibnamefont {Gendiar}},\ }\href {\doibase
  10.1143/PTP.105.409} {\bibfield  {journal} {\bibinfo  {journal} {Prog. Theor.
  Phys.}\ }\textbf {\bibinfo {volume} {105}},\ \bibinfo {pages} {409} (\bibinfo
  {year} {2001})}\BibitemShut {NoStop}%
\bibitem [{\citenamefont {Nishio}\ \emph {et~al.}(2004)\citenamefont {Nishio},
  \citenamefont {Maeshima}, \citenamefont {Gendiar},\ and\ \citenamefont
  {Nishino}}]{nishio2004}%
  \BibitemOpen
  \bibfield  {author} {\bibinfo {author} {\bibfnamefont {Y.}~\bibnamefont
  {Nishio}}, \bibinfo {author} {\bibfnamefont {N.}~\bibnamefont {Maeshima}},
  \bibinfo {author} {\bibfnamefont {A.}~\bibnamefont {Gendiar}}, \ and\
  \bibinfo {author} {\bibfnamefont {T.}~\bibnamefont {Nishino}},\ }\href@noop
  {} {\bibfield  {journal} {\bibinfo  {journal} {Preprint}\ } (\bibinfo {year}
  {2004})},\ \Eprint {http://arxiv.org/abs/cond-mat/0401115}
  {arXiv:cond-mat/0401115} \BibitemShut {NoStop}%
\bibitem [{\citenamefont {Callan}\ and\ \citenamefont
  {Wilczek}(1994)}]{callan_geometric_1994}%
  \BibitemOpen
  \bibfield  {author} {\bibinfo {author} {\bibfnamefont {C.}~\bibnamefont
  {Callan}}\ and\ \bibinfo {author} {\bibfnamefont {F.}~\bibnamefont
  {Wilczek}},\ }\href {\doibase 10.1016/0370-2693(94)91007-3} {\bibfield
  {journal} {\bibinfo  {journal} {arXiv:hep-th/9401072}\ } (\bibinfo {year}
  {1994}),\ 10.1016/0370-2693(94)91007-3},\ \bibinfo {note} {phys.Lett. B333
  (1994) 55-61}\BibitemShut {NoStop}%
\bibitem [{\citenamefont {Calabrese}\ and\ \citenamefont
  {Cardy}(2007)}]{calabrese_entanglement_2007}%
  \BibitemOpen
  \bibfield  {author} {\bibinfo {author} {\bibfnamefont {P.}~\bibnamefont
  {Calabrese}}\ and\ \bibinfo {author} {\bibfnamefont {J.}~\bibnamefont
  {Cardy}},\ }\href {\doibase 10.1088/1742-5468/2007/10/P10004} {\bibfield
  {journal} {\bibinfo  {journal} {J. Stat. Mech.}\ }\textbf {\bibinfo {volume}
  {2007}},\ \bibinfo {pages} {P10004} (\bibinfo {year} {2007})}\BibitemShut
  {NoStop}%
\bibitem [{\citenamefont {Vidal}\ \emph {et~al.}(2003)\citenamefont {Vidal},
  \citenamefont {Latorre}, \citenamefont {Rico},\ and\ \citenamefont
  {Kitaev}}]{vidal_entanglement_2003}%
  \BibitemOpen
  \bibfield  {author} {\bibinfo {author} {\bibfnamefont {G.}~\bibnamefont
  {Vidal}}, \bibinfo {author} {\bibfnamefont {J.~I.}\ \bibnamefont {Latorre}},
  \bibinfo {author} {\bibfnamefont {E.}~\bibnamefont {Rico}}, \ and\ \bibinfo
  {author} {\bibfnamefont {A.}~\bibnamefont {Kitaev}},\ }\href {\doibase
  10.1103/PhysRevLett.90.227902} {\bibfield  {journal} {\bibinfo  {journal}
  {Phys. Rev. Lett.}\ }\textbf {\bibinfo {volume} {90}},\ \bibinfo {pages}
  {227902} (\bibinfo {year} {2003})}\BibitemShut {NoStop}%
\bibitem [{\citenamefont {Tagliacozzo}\ \emph {et~al.}(2008)\citenamefont
  {Tagliacozzo}, \citenamefont {de~Oliveira}, \citenamefont {Iblisdir},\ and\
  \citenamefont {Latorre}}]{tagliacozzo08}%
  \BibitemOpen
  \bibfield  {author} {\bibinfo {author} {\bibfnamefont {L.}~\bibnamefont
  {Tagliacozzo}}, \bibinfo {author} {\bibfnamefont {T.~R.}\ \bibnamefont
  {de~Oliveira}}, \bibinfo {author} {\bibfnamefont {S.}~\bibnamefont
  {Iblisdir}}, \ and\ \bibinfo {author} {\bibfnamefont {J.~I.}\ \bibnamefont
  {Latorre}},\ }\href {\doibase 10.1103/PhysRevB.78.024410} {\bibfield
  {journal} {\bibinfo  {journal} {Phys. Rev. B}\ }\textbf {\bibinfo {volume}
  {78}},\ \bibinfo {pages} {024410} (\bibinfo {year} {2008})}\BibitemShut
  {NoStop}%
\bibitem [{\citenamefont {Pollmann}\ \emph {et~al.}(2009)\citenamefont
  {Pollmann}, \citenamefont {Mukerjee}, \citenamefont {Turner},\ and\
  \citenamefont {Moore}}]{pollmann2009}%
  \BibitemOpen
  \bibfield  {author} {\bibinfo {author} {\bibfnamefont {F.}~\bibnamefont
  {Pollmann}}, \bibinfo {author} {\bibfnamefont {S.}~\bibnamefont {Mukerjee}},
  \bibinfo {author} {\bibfnamefont {A.~M.}\ \bibnamefont {Turner}}, \ and\
  \bibinfo {author} {\bibfnamefont {J.~E.}\ \bibnamefont {Moore}},\ }\href
  {\doibase 10.1103/PhysRevLett.102.255701} {\bibfield  {journal} {\bibinfo
  {journal} {Phys. Rev. Lett.}\ }\textbf {\bibinfo {volume} {102}},\ \bibinfo
  {pages} {255701} (\bibinfo {year} {2009})}\BibitemShut {NoStop}%
\bibitem [{\citenamefont {Pirvu}\ \emph {et~al.}(2012)\citenamefont {Pirvu},
  \citenamefont {Vidal}, \citenamefont {Verstraete},\ and\ \citenamefont
  {Tagliacozzo}}]{pirvu12}%
  \BibitemOpen
  \bibfield  {author} {\bibinfo {author} {\bibfnamefont {B.}~\bibnamefont
  {Pirvu}}, \bibinfo {author} {\bibfnamefont {G.}~\bibnamefont {Vidal}},
  \bibinfo {author} {\bibfnamefont {F.}~\bibnamefont {Verstraete}}, \ and\
  \bibinfo {author} {\bibfnamefont {L.}~\bibnamefont {Tagliacozzo}},\ }\href
  {\doibase 10.1103/PhysRevB.86.075117} {\bibfield  {journal} {\bibinfo
  {journal} {Phys. Rev. B}\ }\textbf {\bibinfo {volume} {86}},\ \bibinfo
  {pages} {075117} (\bibinfo {year} {2012})}\BibitemShut {NoStop}%
\bibitem [{Note1()}]{Note1}%
  \BibitemOpen
  \bibinfo {note} {See however also the recent developments of the analytical
  bootstrap \cite {rychkov_epsilon_2015,
  gliozzi_generalized_2017}.}\BibitemShut {Stop}%
\bibitem [{\citenamefont {Liao}\ \emph {et~al.}(2016)\citenamefont {Liao},
  \citenamefont {Xie}, \citenamefont {Chen}, \citenamefont {Han}, \citenamefont
  {Xie}, \citenamefont {Normand},\ and\ \citenamefont
  {Xiang}}]{liao_heisenberg_2016}%
  \BibitemOpen
  \bibfield  {author} {\bibinfo {author} {\bibfnamefont {H.~J.}\ \bibnamefont
  {Liao}}, \bibinfo {author} {\bibfnamefont {Z.~Y.}\ \bibnamefont {Xie}},
  \bibinfo {author} {\bibfnamefont {J.}~\bibnamefont {Chen}}, \bibinfo {author}
  {\bibfnamefont {X.~J.}\ \bibnamefont {Han}}, \bibinfo {author} {\bibfnamefont
  {H.~D.}\ \bibnamefont {Xie}}, \bibinfo {author} {\bibfnamefont
  {B.}~\bibnamefont {Normand}}, \ and\ \bibinfo {author} {\bibfnamefont
  {T.}~\bibnamefont {Xiang}},\ }\href {\doibase 10.1103/PhysRevB.93.075154}
  {\bibfield  {journal} {\bibinfo  {journal} {Phys. Rev. B}\ }\textbf {\bibinfo
  {volume} {93}},\ \bibinfo {pages} {075154} (\bibinfo {year}
  {2016})}\BibitemShut {NoStop}%
\bibitem [{\citenamefont {Liao}\ \emph
  {et~al.}(2017{\natexlab{a}})\citenamefont {Liao}, \citenamefont {Xie},
  \citenamefont {Chen}, \citenamefont {Liu}, \citenamefont {Xie}, \citenamefont
  {Huang}, \citenamefont {Normand},\ and\ \citenamefont
  {Xiang}}]{liao_gapless_2017}%
  \BibitemOpen
  \bibfield  {author} {\bibinfo {author} {\bibfnamefont {H.~J.}\ \bibnamefont
  {Liao}}, \bibinfo {author} {\bibfnamefont {Z.~Y.}\ \bibnamefont {Xie}},
  \bibinfo {author} {\bibfnamefont {J.}~\bibnamefont {Chen}}, \bibinfo {author}
  {\bibfnamefont {Z.~Y.}\ \bibnamefont {Liu}}, \bibinfo {author} {\bibfnamefont
  {H.~D.}\ \bibnamefont {Xie}}, \bibinfo {author} {\bibfnamefont {R.~Z.}\
  \bibnamefont {Huang}}, \bibinfo {author} {\bibfnamefont {B.}~\bibnamefont
  {Normand}}, \ and\ \bibinfo {author} {\bibfnamefont {T.}~\bibnamefont
  {Xiang}},\ }\href {\doibase 10.1103/PhysRevLett.118.137202} {\bibfield
  {journal} {\bibinfo  {journal} {Phys. Rev. Lett.}\ }\textbf {\bibinfo
  {volume} {118}},\ \bibinfo {pages} {137202} (\bibinfo {year}
  {2017}{\natexlab{a}})}\BibitemShut {NoStop}%
\bibitem [{\citenamefont {Poilblanc}\ and\ \citenamefont
  {Mambrini}(2017)}]{poilblanc_quantum_2017}%
  \BibitemOpen
  \bibfield  {author} {\bibinfo {author} {\bibfnamefont {D.}~\bibnamefont
  {Poilblanc}}\ and\ \bibinfo {author} {\bibfnamefont {M.}~\bibnamefont
  {Mambrini}},\ }\href {\doibase 10.1103/PhysRevB.96.014414} {\bibfield
  {journal} {\bibinfo  {journal} {Phys. Rev. B}\ }\textbf {\bibinfo {volume}
  {96}},\ \bibinfo {pages} {014414} (\bibinfo {year} {2017})}\BibitemShut
  {NoStop}%
\bibitem [{\citenamefont {Verstraete}\ \emph {et~al.}(2006)\citenamefont
  {Verstraete}, \citenamefont {Wolf}, \citenamefont {Perez-Garcia},\ and\
  \citenamefont {Cirac}}]{verstraete_criticality_2006}%
  \BibitemOpen
  \bibfield  {author} {\bibinfo {author} {\bibfnamefont {F.}~\bibnamefont
  {Verstraete}}, \bibinfo {author} {\bibfnamefont {M.~M.}\ \bibnamefont
  {Wolf}}, \bibinfo {author} {\bibfnamefont {D.}~\bibnamefont {Perez-Garcia}},
  \ and\ \bibinfo {author} {\bibfnamefont {J.~I.}\ \bibnamefont {Cirac}},\
  }\href {\doibase 10.1103/PhysRevLett.96.220601} {\bibfield  {journal}
  {\bibinfo  {journal} {Phys. Rev. Lett.}\ }\textbf {\bibinfo {volume} {96}},\
  \bibinfo {pages} {220601} (\bibinfo {year} {2006})}\BibitemShut {NoStop}%
\bibitem [{Note2()}]{Note2}%
  \BibitemOpen
  \bibinfo {note} {E.g. the 2D classical partition function of the critical 2D
  Ising model represented as an iPEPS with $D=2$.}\BibitemShut {Stop}%
\bibitem [{\citenamefont {Henley}(2004)}]{henley_classical_2004}%
  \BibitemOpen
  \bibfield  {author} {\bibinfo {author} {\bibfnamefont {C.~L.}\ \bibnamefont
  {Henley}},\ }\href {\doibase 10.1088/0953-8984/16/11/045} {\bibfield
  {journal} {\bibinfo  {journal} {J. Phys.: Condens. Matter}\ }\textbf
  {\bibinfo {volume} {16}},\ \bibinfo {pages} {S891} (\bibinfo {year}
  {2004})}\BibitemShut {NoStop}%
\bibitem [{\citenamefont {Ardonne}\ \emph {et~al.}(2004)\citenamefont
  {Ardonne}, \citenamefont {Fendley},\ and\ \citenamefont
  {Fradkin}}]{ardonne_topological_2004}%
  \BibitemOpen
  \bibfield  {author} {\bibinfo {author} {\bibfnamefont {E.}~\bibnamefont
  {Ardonne}}, \bibinfo {author} {\bibfnamefont {P.}~\bibnamefont {Fendley}}, \
  and\ \bibinfo {author} {\bibfnamefont {E.}~\bibnamefont {Fradkin}},\ }\href
  {\doibase 10.1016/j.aop.2004.01.004} {\bibfield  {journal} {\bibinfo
  {journal} {Annals of Physics}\ }\textbf {\bibinfo {volume} {310}},\ \bibinfo
  {pages} {493} (\bibinfo {year} {2004})}\BibitemShut {NoStop}%
\bibitem [{\citenamefont {Castelnovo}\ \emph {et~al.}(2005)\citenamefont
  {Castelnovo}, \citenamefont {Chamon}, \citenamefont {Mudry},\ and\
  \citenamefont {Pujol}}]{castelnovo_quantum_2005}%
  \BibitemOpen
  \bibfield  {author} {\bibinfo {author} {\bibfnamefont {C.}~\bibnamefont
  {Castelnovo}}, \bibinfo {author} {\bibfnamefont {C.}~\bibnamefont {Chamon}},
  \bibinfo {author} {\bibfnamefont {C.}~\bibnamefont {Mudry}}, \ and\ \bibinfo
  {author} {\bibfnamefont {P.}~\bibnamefont {Pujol}},\ }\href {\doibase
  10.1016/j.aop.2005.01.006} {\bibfield  {journal} {\bibinfo  {journal} {Annals
  of Physics}\ }\textbf {\bibinfo {volume} {318}},\ \bibinfo {pages} {316}
  (\bibinfo {year} {2005})}\BibitemShut {NoStop}%
\bibitem [{\citenamefont {Isakov}\ \emph {et~al.}(2011)\citenamefont {Isakov},
  \citenamefont {Fendley}, \citenamefont {Ludwig}, \citenamefont {Trebst},\
  and\ \citenamefont {Troyer}}]{isakov_dynamics_2011}%
  \BibitemOpen
  \bibfield  {author} {\bibinfo {author} {\bibfnamefont {S.~V.}\ \bibnamefont
  {Isakov}}, \bibinfo {author} {\bibfnamefont {P.}~\bibnamefont {Fendley}},
  \bibinfo {author} {\bibfnamefont {A.~W.~W.}\ \bibnamefont {Ludwig}}, \bibinfo
  {author} {\bibfnamefont {S.}~\bibnamefont {Trebst}}, \ and\ \bibinfo {author}
  {\bibfnamefont {M.}~\bibnamefont {Troyer}},\ }\href {\doibase
  10.1103/PhysRevB.83.125114} {\bibfield  {journal} {\bibinfo  {journal} {Phys.
  Rev. B}\ }\textbf {\bibinfo {volume} {83}},\ \bibinfo {pages} {125114}
  (\bibinfo {year} {2011})}\BibitemShut {NoStop}%
\bibitem [{\citenamefont {Tagliacozzo}\ \emph {et~al.}(2014)\citenamefont
  {Tagliacozzo}, \citenamefont {Celi},\ and\ \citenamefont
  {Lewenstein}}]{tagliacozzo_tensor_2014}%
  \BibitemOpen
  \bibfield  {author} {\bibinfo {author} {\bibfnamefont {L.}~\bibnamefont
  {Tagliacozzo}}, \bibinfo {author} {\bibfnamefont {A.}~\bibnamefont {Celi}}, \
  and\ \bibinfo {author} {\bibfnamefont {M.}~\bibnamefont {Lewenstein}},\
  }\href {\doibase 10.1103/PhysRevX.4.041024} {\bibfield  {journal} {\bibinfo
  {journal} {Phys. Rev. X}\ }\textbf {\bibinfo {volume} {4}},\ \bibinfo {pages}
  {041024} (\bibinfo {year} {2014})}\BibitemShut {NoStop}%
\bibitem [{\citenamefont {Zohar}\ \emph {et~al.}(2015)\citenamefont {Zohar},
  \citenamefont {Burrello}, \citenamefont {Wahl},\ and\ \citenamefont
  {Cirac}}]{zohar_fermionic_2015}%
  \BibitemOpen
  \bibfield  {author} {\bibinfo {author} {\bibfnamefont {E.}~\bibnamefont
  {Zohar}}, \bibinfo {author} {\bibfnamefont {M.}~\bibnamefont {Burrello}},
  \bibinfo {author} {\bibfnamefont {T.~B.}\ \bibnamefont {Wahl}}, \ and\
  \bibinfo {author} {\bibfnamefont {J.~I.}\ \bibnamefont {Cirac}},\ }\href
  {\doibase 10.1016/j.aop.2015.10.009} {\bibfield  {journal} {\bibinfo
  {journal} {Annals of Physics}\ }\textbf {\bibinfo {volume} {363}},\ \bibinfo
  {pages} {385} (\bibinfo {year} {2015})}\BibitemShut {NoStop}%
\bibitem [{\citenamefont {Zohar}\ \emph {et~al.}(2016)\citenamefont {Zohar},
  \citenamefont {Wahl}, \citenamefont {Burrello},\ and\ \citenamefont
  {Cirac}}]{zohar_projected_2016}%
  \BibitemOpen
  \bibfield  {author} {\bibinfo {author} {\bibfnamefont {E.}~\bibnamefont
  {Zohar}}, \bibinfo {author} {\bibfnamefont {T.~B.}\ \bibnamefont {Wahl}},
  \bibinfo {author} {\bibfnamefont {M.}~\bibnamefont {Burrello}}, \ and\
  \bibinfo {author} {\bibfnamefont {J.~I.}\ \bibnamefont {Cirac}},\ }\href
  {\doibase 10.1016/j.aop.2016.08.008} {\bibfield  {journal} {\bibinfo
  {journal} {Annals of Physics}\ }\textbf {\bibinfo {volume} {374}},\ \bibinfo
  {pages} {84} (\bibinfo {year} {2016})}\BibitemShut {NoStop}%
\bibitem [{\citenamefont {Ge}\ and\ \citenamefont
  {Eisert}(2016)}]{ge_area_2016}%
  \BibitemOpen
  \bibfield  {author} {\bibinfo {author} {\bibfnamefont {Y.}~\bibnamefont
  {Ge}}\ and\ \bibinfo {author} {\bibfnamefont {J.}~\bibnamefont {Eisert}},\
  }\href {\doibase 10.1088/1367-2630/18/8/083026} {\bibfield  {journal}
  {\bibinfo  {journal} {New J. Phys.}\ }\textbf {\bibinfo {volume} {18}},\
  \bibinfo {pages} {083026} (\bibinfo {year} {2016})}\BibitemShut {NoStop}%
\bibitem [{\citenamefont {Zhao}\ \emph {et~al.}(2012)\citenamefont {Zhao},
  \citenamefont {Xu}, \citenamefont {Chen}, \citenamefont {Wei}, \citenamefont
  {Qin}, \citenamefont {Zhang},\ and\ \citenamefont {Xiang}}]{Zhao12}%
  \BibitemOpen
  \bibfield  {author} {\bibinfo {author} {\bibfnamefont {H.~H.}\ \bibnamefont
  {Zhao}}, \bibinfo {author} {\bibfnamefont {C.}~\bibnamefont {Xu}}, \bibinfo
  {author} {\bibfnamefont {Q.~N.}\ \bibnamefont {Chen}}, \bibinfo {author}
  {\bibfnamefont {Z.~C.}\ \bibnamefont {Wei}}, \bibinfo {author} {\bibfnamefont
  {M.~P.}\ \bibnamefont {Qin}}, \bibinfo {author} {\bibfnamefont {G.~M.}\
  \bibnamefont {Zhang}}, \ and\ \bibinfo {author} {\bibfnamefont
  {T.}~\bibnamefont {Xiang}},\ }\href {\doibase 10.1103/PhysRevB.85.134416}
  {\bibfield  {journal} {\bibinfo  {journal} {Phys. Rev. B}\ }\textbf {\bibinfo
  {volume} {85}},\ \bibinfo {pages} {134416} (\bibinfo {year}
  {2012})}\BibitemShut {NoStop}%
\bibitem [{\citenamefont {Corboz}\ \emph {et~al.}(2012)\citenamefont {Corboz},
  \citenamefont {Lajk{\'o}}, \citenamefont {L{\"a}uchli}, \citenamefont
  {Penc},\ and\ \citenamefont {Mila}}]{Corboz12_su4}%
  \BibitemOpen
  \bibfield  {author} {\bibinfo {author} {\bibfnamefont {P.}~\bibnamefont
  {Corboz}}, \bibinfo {author} {\bibfnamefont {M.}~\bibnamefont {Lajk{\'o}}},
  \bibinfo {author} {\bibfnamefont {A.~M.}\ \bibnamefont {L{\"a}uchli}},
  \bibinfo {author} {\bibfnamefont {K.}~\bibnamefont {Penc}}, \ and\ \bibinfo
  {author} {\bibfnamefont {F.}~\bibnamefont {Mila}},\ }\href {\doibase
  10.1103/PhysRevX.2.041013} {\bibfield  {journal} {\bibinfo  {journal} {Phys.
  Rev. X}\ }\textbf {\bibinfo {volume} {2}},\ \bibinfo {pages} {041013}
  (\bibinfo {year} {2012})}\BibitemShut {NoStop}%
\bibitem [{\citenamefont {Xie}\ \emph {et~al.}(2014)\citenamefont {Xie},
  \citenamefont {Chen}, \citenamefont {Yu}, \citenamefont {Kong}, \citenamefont
  {Normand},\ and\ \citenamefont {Xiang}}]{xie14}%
  \BibitemOpen
  \bibfield  {author} {\bibinfo {author} {\bibfnamefont {Z.}~\bibnamefont
  {Xie}}, \bibinfo {author} {\bibfnamefont {J.}~\bibnamefont {Chen}}, \bibinfo
  {author} {\bibfnamefont {J.}~\bibnamefont {Yu}}, \bibinfo {author}
  {\bibfnamefont {X.}~\bibnamefont {Kong}}, \bibinfo {author} {\bibfnamefont
  {B.}~\bibnamefont {Normand}}, \ and\ \bibinfo {author} {\bibfnamefont
  {T.}~\bibnamefont {Xiang}},\ }\href {\doibase 10.1103/PhysRevX.4.011025}
  {\bibfield  {journal} {\bibinfo  {journal} {Phys. Rev. X}\ }\textbf {\bibinfo
  {volume} {4}},\ \bibinfo {pages} {011025} (\bibinfo {year}
  {2014})}\BibitemShut {NoStop}%
\bibitem [{\citenamefont {Corboz}\ and\ \citenamefont
  {Mila}(2013)}]{Corboz13_shastry}%
  \BibitemOpen
  \bibfield  {author} {\bibinfo {author} {\bibfnamefont {P.}~\bibnamefont
  {Corboz}}\ and\ \bibinfo {author} {\bibfnamefont {F.}~\bibnamefont {Mila}},\
  }\href {\doibase 10.1103/PhysRevB.87.115144} {\bibfield  {journal} {\bibinfo
  {journal} {Phys. Rev. B}\ }\textbf {\bibinfo {volume} {87}},\ \bibinfo
  {pages} {115144} (\bibinfo {year} {2013})}\BibitemShut {NoStop}%
\bibitem [{\citenamefont {Gu}\ \emph {et~al.}(2013)\citenamefont {Gu},
  \citenamefont {Jiang}, \citenamefont {Sheng}, \citenamefont {Yao},
  \citenamefont {Balents},\ and\ \citenamefont {Wen}}]{gu2013}%
  \BibitemOpen
  \bibfield  {author} {\bibinfo {author} {\bibfnamefont {Z.-C.}\ \bibnamefont
  {Gu}}, \bibinfo {author} {\bibfnamefont {H.-C.}\ \bibnamefont {Jiang}},
  \bibinfo {author} {\bibfnamefont {D.~N.}\ \bibnamefont {Sheng}}, \bibinfo
  {author} {\bibfnamefont {H.}~\bibnamefont {Yao}}, \bibinfo {author}
  {\bibfnamefont {L.}~\bibnamefont {Balents}}, \ and\ \bibinfo {author}
  {\bibfnamefont {X.-G.}\ \bibnamefont {Wen}},\ }\href {\doibase
  10.1103/PhysRevB.88.155112} {\bibfield  {journal} {\bibinfo  {journal} {Phys.
  Rev. B}\ }\textbf {\bibinfo {volume} {88}},\ \bibinfo {pages} {155112}
  (\bibinfo {year} {2013})}\BibitemShut {NoStop}%
\bibitem [{\citenamefont {Corboz}\ \emph {et~al.}(2014)\citenamefont {Corboz},
  \citenamefont {Rice},\ and\ \citenamefont {Troyer}}]{corboz14_tJ}%
  \BibitemOpen
  \bibfield  {author} {\bibinfo {author} {\bibfnamefont {P.}~\bibnamefont
  {Corboz}}, \bibinfo {author} {\bibfnamefont {T.}~\bibnamefont {Rice}}, \ and\
  \bibinfo {author} {\bibfnamefont {M.}~\bibnamefont {Troyer}},\ }\href
  {\doibase 10.1103/PhysRevLett.113.046402} {\bibfield  {journal} {\bibinfo
  {journal} {Phys. Rev. Lett.}\ }\textbf {\bibinfo {volume} {113}},\ \bibinfo
  {pages} {046402} (\bibinfo {year} {2014})}\BibitemShut {NoStop}%
\bibitem [{\citenamefont {Picot}\ and\ \citenamefont
  {Poilblanc}(2015)}]{picot15a}%
  \BibitemOpen
  \bibfield  {author} {\bibinfo {author} {\bibfnamefont {T.}~\bibnamefont
  {Picot}}\ and\ \bibinfo {author} {\bibfnamefont {D.}~\bibnamefont
  {Poilblanc}},\ }\href {\doibase 10.1103/PhysRevB.91.064415} {\bibfield
  {journal} {\bibinfo  {journal} {Phys. Rev. B}\ }\textbf {\bibinfo {volume}
  {91}},\ \bibinfo {pages} {064415} (\bibinfo {year} {2015})}\BibitemShut
  {NoStop}%
\bibitem [{\citenamefont {Picot}\ \emph {et~al.}(2016)\citenamefont {Picot},
  \citenamefont {Ziegler}, \citenamefont {Or\'us},\ and\ \citenamefont
  {Poilblanc}}]{picot15}%
  \BibitemOpen
  \bibfield  {author} {\bibinfo {author} {\bibfnamefont {T.}~\bibnamefont
  {Picot}}, \bibinfo {author} {\bibfnamefont {M.}~\bibnamefont {Ziegler}},
  \bibinfo {author} {\bibfnamefont {R.}~\bibnamefont {Or\'us}}, \ and\ \bibinfo
  {author} {\bibfnamefont {D.}~\bibnamefont {Poilblanc}},\ }\href {\doibase
  10.1103/PhysRevB.93.060407} {\bibfield  {journal} {\bibinfo  {journal} {Phys.
  Rev. B}\ }\textbf {\bibinfo {volume} {93}},\ \bibinfo {pages} {060407}
  (\bibinfo {year} {2016})}\BibitemShut {NoStop}%
\bibitem [{\citenamefont {Liao}\ \emph
  {et~al.}(2017{\natexlab{b}})\citenamefont {Liao}, \citenamefont {Xie},
  \citenamefont {Chen}, \citenamefont {Liu}, \citenamefont {Xie}, \citenamefont
  {Huang}, \citenamefont {Normand},\ and\ \citenamefont {Xiang}}]{liao16}%
  \BibitemOpen
  \bibfield  {author} {\bibinfo {author} {\bibfnamefont {H.~J.}\ \bibnamefont
  {Liao}}, \bibinfo {author} {\bibfnamefont {Z.~Y.}\ \bibnamefont {Xie}},
  \bibinfo {author} {\bibfnamefont {J.}~\bibnamefont {Chen}}, \bibinfo {author}
  {\bibfnamefont {Z.~Y.}\ \bibnamefont {Liu}}, \bibinfo {author} {\bibfnamefont
  {H.~D.}\ \bibnamefont {Xie}}, \bibinfo {author} {\bibfnamefont {R.~Z.}\
  \bibnamefont {Huang}}, \bibinfo {author} {\bibfnamefont {B.}~\bibnamefont
  {Normand}}, \ and\ \bibinfo {author} {\bibfnamefont {T.}~\bibnamefont
  {Xiang}},\ }\href {\doibase 10.1103/PhysRevLett.118.137202} {\bibfield
  {journal} {\bibinfo  {journal} {Phys. Rev. Lett.}\ }\textbf {\bibinfo
  {volume} {118}},\ \bibinfo {pages} {137202} (\bibinfo {year}
  {2017}{\natexlab{b}})}\BibitemShut {NoStop}%
\bibitem [{\citenamefont {Niesen}\ and\ \citenamefont
  {Corboz}(2017)}]{niesen17}%
  \BibitemOpen
  \bibfield  {author} {\bibinfo {author} {\bibfnamefont {I.}~\bibnamefont
  {Niesen}}\ and\ \bibinfo {author} {\bibfnamefont {P.}~\bibnamefont
  {Corboz}},\ }\href {\doibase 10.1103/PhysRevB.95.180404} {\bibfield
  {journal} {\bibinfo  {journal} {Phys. Rev. B}\ }\textbf {\bibinfo {volume}
  {95}},\ \bibinfo {pages} {180404} (\bibinfo {year} {2017})}\BibitemShut
  {NoStop}%
\bibitem [{\citenamefont {Zheng}\ \emph {et~al.}(2017)\citenamefont {Zheng},
  \citenamefont {Chung}, \citenamefont {Corboz}, \citenamefont {Ehlers},
  \citenamefont {Qin}, \citenamefont {Noack}, \citenamefont {Shi},
  \citenamefont {White}, \citenamefont {Zhang},\ and\ \citenamefont
  {Chan}}]{zheng17}%
  \BibitemOpen
  \bibfield  {author} {\bibinfo {author} {\bibfnamefont {B.-X.}\ \bibnamefont
  {Zheng}}, \bibinfo {author} {\bibfnamefont {C.-M.}\ \bibnamefont {Chung}},
  \bibinfo {author} {\bibfnamefont {P.}~\bibnamefont {Corboz}}, \bibinfo
  {author} {\bibfnamefont {G.}~\bibnamefont {Ehlers}}, \bibinfo {author}
  {\bibfnamefont {M.-P.}\ \bibnamefont {Qin}}, \bibinfo {author} {\bibfnamefont
  {R.~M.}\ \bibnamefont {Noack}}, \bibinfo {author} {\bibfnamefont
  {H.}~\bibnamefont {Shi}}, \bibinfo {author} {\bibfnamefont {S.~R.}\
  \bibnamefont {White}}, \bibinfo {author} {\bibfnamefont {S.}~\bibnamefont
  {Zhang}}, \ and\ \bibinfo {author} {\bibfnamefont {G.~K.-L.}\ \bibnamefont
  {Chan}},\ }\href {\doibase 10.1126/science.aam7127} {\bibfield  {journal}
  {\bibinfo  {journal} {Science}\ }\textbf {\bibinfo {volume} {358}},\ \bibinfo
  {pages} {1155} (\bibinfo {year} {2017})}\BibitemShut {NoStop}%
\bibitem [{\citenamefont {Haghshenas}\ and\ \citenamefont
  {Sheng}(2017)}]{haghshenas17}%
  \BibitemOpen
  \bibfield  {author} {\bibinfo {author} {\bibfnamefont {R.}~\bibnamefont
  {Haghshenas}}\ and\ \bibinfo {author} {\bibfnamefont {D.~N.}\ \bibnamefont
  {Sheng}},\ }\href {http://arxiv.org/abs/1711.07584} {\bibfield  {journal}
  {\bibinfo  {journal} {arXiv:1711.07584 [cond-mat, physics:quant-ph]}\ }
  (\bibinfo {year} {2017})},\ \bibinfo {note} {arXiv: 1711.07584}\BibitemShut
  {NoStop}%
\bibitem [{\citenamefont {Corboz}\ \emph {et~al.}(2010)\citenamefont {Corboz},
  \citenamefont {Orus}, \citenamefont {Bauer},\ and\ \citenamefont
  {Vidal}}]{corboz2010}%
  \BibitemOpen
  \bibfield  {author} {\bibinfo {author} {\bibfnamefont {P.}~\bibnamefont
  {Corboz}}, \bibinfo {author} {\bibfnamefont {R.}~\bibnamefont {Orus}},
  \bibinfo {author} {\bibfnamefont {B.}~\bibnamefont {Bauer}}, \ and\ \bibinfo
  {author} {\bibfnamefont {G.}~\bibnamefont {Vidal}},\ }\href {\doibase
  10.1103/PhysRevB.81.165104} {\bibfield  {journal} {\bibinfo  {journal} {Phys.
  Rev. B}\ }\textbf {\bibinfo {volume} {81}},\ \bibinfo {pages} {165104}
  (\bibinfo {year} {2010})}\BibitemShut {NoStop}%
\bibitem [{\citenamefont {Phien}\ \emph {et~al.}(2015)\citenamefont {Phien},
  \citenamefont {Bengua}, \citenamefont {Tuan}, \citenamefont {Corboz},\ and\
  \citenamefont {Orus}}]{phien15}%
  \BibitemOpen
  \bibfield  {author} {\bibinfo {author} {\bibfnamefont {H.~N.}\ \bibnamefont
  {Phien}}, \bibinfo {author} {\bibfnamefont {J.~A.}\ \bibnamefont {Bengua}},
  \bibinfo {author} {\bibfnamefont {H.~D.}\ \bibnamefont {Tuan}}, \bibinfo
  {author} {\bibfnamefont {P.}~\bibnamefont {Corboz}}, \ and\ \bibinfo {author}
  {\bibfnamefont {R.}~\bibnamefont {Orus}},\ }\href {\doibase
  10.1103/PhysRevB.92.035142} {\bibfield  {journal} {\bibinfo  {journal} {Phys.
  Rev. B}\ }\textbf {\bibinfo {volume} {92}},\ \bibinfo {pages} {035142}
  (\bibinfo {year} {2015})}\BibitemShut {NoStop}%
\bibitem [{\citenamefont {Corboz}(2016)}]{corboz16b}%
  \BibitemOpen
  \bibfield  {author} {\bibinfo {author} {\bibfnamefont {P.}~\bibnamefont
  {Corboz}},\ }\href {\doibase 10.1103/PhysRevB.94.035133} {\bibfield
  {journal} {\bibinfo  {journal} {Phys. Rev. B}\ }\textbf {\bibinfo {volume}
  {94}},\ \bibinfo {pages} {035133} (\bibinfo {year} {2016})}\BibitemShut
  {NoStop}%
\bibitem [{\citenamefont {Vanderstraeten}\ \emph {et~al.}(2016)\citenamefont
  {Vanderstraeten}, \citenamefont {Haegeman}, \citenamefont {Corboz},\ and\
  \citenamefont {Verstraete}}]{vanderstraeten16}%
  \BibitemOpen
  \bibfield  {author} {\bibinfo {author} {\bibfnamefont {L.}~\bibnamefont
  {Vanderstraeten}}, \bibinfo {author} {\bibfnamefont {J.}~\bibnamefont
  {Haegeman}}, \bibinfo {author} {\bibfnamefont {P.}~\bibnamefont {Corboz}}, \
  and\ \bibinfo {author} {\bibfnamefont {F.}~\bibnamefont {Verstraete}},\
  }\href {\doibase 10.1103/PhysRevB.94.155123} {\bibfield  {journal} {\bibinfo
  {journal} {Phys. Rev. B}\ }\textbf {\bibinfo {volume} {94}},\ \bibinfo
  {pages} {155123} (\bibinfo {year} {2016})}\BibitemShut {NoStop}%
\bibitem [{\citenamefont {Nishino}\ and\ \citenamefont
  {Okunishi}(1996)}]{nishino1996}%
  \BibitemOpen
  \bibfield  {author} {\bibinfo {author} {\bibfnamefont {T.}~\bibnamefont
  {Nishino}}\ and\ \bibinfo {author} {\bibfnamefont {K.}~\bibnamefont
  {Okunishi}},\ }\href {\doibase 10.1143/JPSJ.65.891} {\bibfield  {journal}
  {\bibinfo  {journal} {J. Phys. Soc. Jpn.}\ }\textbf {\bibinfo {volume}
  {65}},\ \bibinfo {pages} {891} (\bibinfo {year} {1996})}\BibitemShut
  {NoStop}%
\bibitem [{\citenamefont {Or\'us}\ and\ \citenamefont
  {Vidal}(2009)}]{orus2009-1}%
  \BibitemOpen
  \bibfield  {author} {\bibinfo {author} {\bibfnamefont {R.}~\bibnamefont
  {Or\'us}}\ and\ \bibinfo {author} {\bibfnamefont {G.}~\bibnamefont {Vidal}},\
  }\href {\doibase 10.1103/PhysRevB.80.094403} {\bibfield  {journal} {\bibinfo
  {journal} {Phys. Rev. B}\ }\textbf {\bibinfo {volume} {80}},\ \bibinfo
  {pages} {094403} (\bibinfo {year} {2009})}\BibitemShut {NoStop}%
\bibitem [{\citenamefont {Singh}\ \emph {et~al.}(2011)\citenamefont {Singh},
  \citenamefont {Pfeifer},\ and\ \citenamefont {Vidal}}]{singh2010}%
  \BibitemOpen
  \bibfield  {author} {\bibinfo {author} {\bibfnamefont {S.}~\bibnamefont
  {Singh}}, \bibinfo {author} {\bibfnamefont {R.~N.~C.}\ \bibnamefont
  {Pfeifer}}, \ and\ \bibinfo {author} {\bibfnamefont {G.}~\bibnamefont
  {Vidal}},\ }\href {\doibase 10.1103/PhysRevB.83.115125} {\bibfield  {journal}
  {\bibinfo  {journal} {Phys. Rev. B}\ }\textbf {\bibinfo {volume} {83}},\
  \bibinfo {pages} {115125} (\bibinfo {year} {2011})}\BibitemShut {NoStop}%
\bibitem [{\citenamefont {Bauer}\ \emph {et~al.}(2011)\citenamefont {Bauer},
  \citenamefont {Corboz}, \citenamefont {Or\'us},\ and\ \citenamefont
  {Troyer}}]{bauer2011}%
  \BibitemOpen
  \bibfield  {author} {\bibinfo {author} {\bibfnamefont {B.}~\bibnamefont
  {Bauer}}, \bibinfo {author} {\bibfnamefont {P.}~\bibnamefont {Corboz}},
  \bibinfo {author} {\bibfnamefont {R.}~\bibnamefont {Or\'us}}, \ and\ \bibinfo
  {author} {\bibfnamefont {M.}~\bibnamefont {Troyer}},\ }\href {\doibase
  10.1103/PhysRevB.83.125106} {\bibfield  {journal} {\bibinfo  {journal} {Phys.
  Rev. B}\ }\textbf {\bibinfo {volume} {83}},\ \bibinfo {pages} {125106}
  (\bibinfo {year} {2011})}\BibitemShut {NoStop}%
\bibitem [{Note3()}]{Note3}%
  \BibitemOpen
  \bibinfo {note} {If $H$ is gapless, a projector onto a well defined state can
  be obtained by first opening an infinitesimally small gap of order $\epsilon
  $ modifying $H \to H_{\epsilon }$, and then taking the limit $\beta \to
  \infty $. We will avoid similar details in the rest of the
  section.}\BibitemShut {Stop}%
\bibitem [{\citenamefont {Vidal}(2007)}]{vidal_classical_2007}%
  \BibitemOpen
  \bibfield  {author} {\bibinfo {author} {\bibfnamefont {G.}~\bibnamefont
  {Vidal}},\ }\href {\doibase 10.1103/PhysRevLett.98.070201} {\bibfield
  {journal} {\bibinfo  {journal} {Phys. Rev. Lett.}\ }\textbf {\bibinfo
  {volume} {98}},\ \bibinfo {pages} {070201} (\bibinfo {year}
  {2007})}\BibitemShut {NoStop}%
\bibitem [{\citenamefont {Vidal}(2004)}]{vidal_efficient_2004}%
  \BibitemOpen
  \bibfield  {author} {\bibinfo {author} {\bibfnamefont {G.}~\bibnamefont
  {Vidal}},\ }\href {\doibase 10.1103/PhysRevLett.93.040502} {\bibfield
  {journal} {\bibinfo  {journal} {Phys. Rev. Lett.}\ }\textbf {\bibinfo
  {volume} {93}},\ \bibinfo {pages} {040502} (\bibinfo {year}
  {2004})}\BibitemShut {NoStop}%
\bibitem [{\citenamefont {Or{\'u}s}\ and\ \citenamefont
  {Vidal}(2008)}]{orus_infinite_2008}%
  \BibitemOpen
  \bibfield  {author} {\bibinfo {author} {\bibfnamefont {R.}~\bibnamefont
  {Or{\'u}s}}\ and\ \bibinfo {author} {\bibfnamefont {G.}~\bibnamefont
  {Vidal}},\ }\href {\doibase 10.1103/PhysRevB.78.155117} {\bibfield  {journal}
  {\bibinfo  {journal} {Phys. Rev. B}\ }\textbf {\bibinfo {volume} {78}},\
  \bibinfo {pages} {155117} (\bibinfo {year} {2008})}\BibitemShut {NoStop}%
\bibitem [{\citenamefont {Ferris}(2015)}]{ferris15}%
  \BibitemOpen
  \bibfield  {author} {\bibinfo {author} {\bibfnamefont {A.~J.}\ \bibnamefont
  {Ferris}},\ }\href {http://arxiv.org/abs/1507.00767} {\bibfield  {journal}
  {\bibinfo  {journal} {arXiv:1507.00767}\ } (\bibinfo {year}
  {2015})}\BibitemShut {NoStop}%
\bibitem [{\citenamefont {Cardy}(1996)}]{cardy_scaling_1996}%
  \BibitemOpen
  \bibfield  {author} {\bibinfo {author} {\bibfnamefont {J.}~\bibnamefont
  {Cardy}},\ }\href@noop {} {\emph {\bibinfo {title} {Scaling and
  {Renormalization} in {Statistical} {Physics}}}}\ (\bibinfo  {publisher}
  {Cambridge University Press},\ \bibinfo {year} {1996})\BibitemShut {NoStop}%
\bibitem [{\citenamefont {Hastings}\ and\ \citenamefont
  {Koma}(2006)}]{hastings_spectral_2006}%
  \BibitemOpen
  \bibfield  {author} {\bibinfo {author} {\bibfnamefont {M.~B.}\ \bibnamefont
  {Hastings}}\ and\ \bibinfo {author} {\bibfnamefont {T.}~\bibnamefont
  {Koma}},\ }\href {\doibase 10.1007/s00220-006-0030-4} {\bibfield  {journal}
  {\bibinfo  {journal} {Commun. Math. Phys.}\ }\textbf {\bibinfo {volume}
  {265}},\ \bibinfo {pages} {781} (\bibinfo {year} {2006})}\BibitemShut
  {NoStop}%
\bibitem [{\citenamefont {Nishino}\ \emph {et~al.}(1996)\citenamefont
  {Nishino}, \citenamefont {Okunishi},\ and\ \citenamefont
  {Kikuchi}}]{nishino96}%
  \BibitemOpen
  \bibfield  {author} {\bibinfo {author} {\bibfnamefont {T.}~\bibnamefont
  {Nishino}}, \bibinfo {author} {\bibfnamefont {K.}~\bibnamefont {Okunishi}}, \
  and\ \bibinfo {author} {\bibfnamefont {M.}~\bibnamefont {Kikuchi}},\ }\href
  {\doibase 10.1016/0375-9601(96)00128-4} {\bibfield  {journal} {\bibinfo
  {journal} {Phys. Lett. A}\ }\textbf {\bibinfo {volume} {213}},\ \bibinfo
  {pages} {69} (\bibinfo {year} {1996})}\BibitemShut {NoStop}%
\bibitem [{\citenamefont {Wang}\ \emph {et~al.}(2014)\citenamefont {Wang},
  \citenamefont {Corboz},\ and\ \citenamefont {Troyer}}]{wang14}%
  \BibitemOpen
  \bibfield  {author} {\bibinfo {author} {\bibfnamefont {L.}~\bibnamefont
  {Wang}}, \bibinfo {author} {\bibfnamefont {P.}~\bibnamefont {Corboz}}, \ and\
  \bibinfo {author} {\bibfnamefont {M.}~\bibnamefont {Troyer}},\ }\href
  {\doibase 10.1088/1367-2630/16/10/103008} {\bibfield  {journal} {\bibinfo
  {journal} {New J. Phys.}\ }\textbf {\bibinfo {volume} {16}},\ \bibinfo
  {pages} {103008} (\bibinfo {year} {2014})}\BibitemShut {NoStop}%
\bibitem [{\citenamefont {Scherer}\ \emph {et~al.}(2015)\citenamefont
  {Scherer}, \citenamefont {Scherer},\ and\ \citenamefont
  {Honerkamp}}]{scherer15}%
  \BibitemOpen
  \bibfield  {author} {\bibinfo {author} {\bibfnamefont {D.~D.}\ \bibnamefont
  {Scherer}}, \bibinfo {author} {\bibfnamefont {M.~M.}\ \bibnamefont
  {Scherer}}, \ and\ \bibinfo {author} {\bibfnamefont {C.}~\bibnamefont
  {Honerkamp}},\ }\href {\doibase 10.1103/PhysRevB.92.155137} {\bibfield
  {journal} {\bibinfo  {journal} {Phys. Rev. B}\ }\textbf {\bibinfo {volume}
  {92}},\ \bibinfo {pages} {155137} (\bibinfo {year} {2015})}\BibitemShut
  {NoStop}%
\bibitem [{\citenamefont {Li}\ \emph {et~al.}(2015)\citenamefont {Li},
  \citenamefont {Jiang},\ and\ \citenamefont {Yao}}]{li15}%
  \BibitemOpen
  \bibfield  {author} {\bibinfo {author} {\bibfnamefont {Z.-X.}\ \bibnamefont
  {Li}}, \bibinfo {author} {\bibfnamefont {Y.-F.}\ \bibnamefont {Jiang}}, \
  and\ \bibinfo {author} {\bibfnamefont {H.}~\bibnamefont {Yao}},\ }\href
  {\doibase 10.1088/1367-2630/17/8/085003} {\bibfield  {journal} {\bibinfo
  {journal} {New J. Phys.}\ }\textbf {\bibinfo {volume} {17}},\ \bibinfo
  {pages} {085003} (\bibinfo {year} {2015})}\BibitemShut {NoStop}%
\bibitem [{\citenamefont {Wang}\ \emph {et~al.}(2015)\citenamefont {Wang},
  \citenamefont {Iazzi}, \citenamefont {Corboz},\ and\ \citenamefont
  {Troyer}}]{wang15b}%
  \BibitemOpen
  \bibfield  {author} {\bibinfo {author} {\bibfnamefont {L.}~\bibnamefont
  {Wang}}, \bibinfo {author} {\bibfnamefont {M.}~\bibnamefont {Iazzi}},
  \bibinfo {author} {\bibfnamefont {P.}~\bibnamefont {Corboz}}, \ and\ \bibinfo
  {author} {\bibfnamefont {M.}~\bibnamefont {Troyer}},\ }\href {\doibase
  10.1103/PhysRevB.91.235151} {\bibfield  {journal} {\bibinfo  {journal} {Phys.
  Rev. B}\ }\textbf {\bibinfo {volume} {91}},\ \bibinfo {pages} {235151}
  (\bibinfo {year} {2015})}\BibitemShut {NoStop}%
\bibitem [{\citenamefont {Wang}\ \emph {et~al.}(2016)\citenamefont {Wang},
  \citenamefont {Liu},\ and\ \citenamefont {Troyer}}]{wang16b}%
  \BibitemOpen
  \bibfield  {author} {\bibinfo {author} {\bibfnamefont {L.}~\bibnamefont
  {Wang}}, \bibinfo {author} {\bibfnamefont {Y.-H.}\ \bibnamefont {Liu}}, \
  and\ \bibinfo {author} {\bibfnamefont {M.}~\bibnamefont {Troyer}},\ }\href
  {\doibase 10.1103/PhysRevB.93.155117} {\bibfield  {journal} {\bibinfo
  {journal} {Phys. Rev. B}\ }\textbf {\bibinfo {volume} {93}},\ \bibinfo
  {pages} {155117} (\bibinfo {year} {2016})}\BibitemShut {NoStop}%
\bibitem [{\citenamefont {Capponi}(2017)}]{capponi17}%
  \BibitemOpen
  \bibfield  {author} {\bibinfo {author} {\bibfnamefont {S.}~\bibnamefont
  {Capponi}},\ }\href {\doibase 10.1088/1361-648X/29/4/043002} {\bibfield
  {journal} {\bibinfo  {journal} {J. Phys.: Condens. Matter}\ }\textbf
  {\bibinfo {volume} {29}},\ \bibinfo {pages} {043002} (\bibinfo {year}
  {2017})}\BibitemShut {NoStop}%
\bibitem [{Note4()}]{Note4}%
  \BibitemOpen
  \bibinfo {note} {The correlation length is computed from the largest and
  second largest eigenvalues of the row-to-row transfer matrix as in Ref.~\cite
  {nishino96}}\BibitemShut {NoStop}%
\bibitem [{Note5()}]{Note5}%
  \BibitemOpen
  \bibinfo {note} {During completion of this work an interesting alternative
  approach to obtain the correlation length in the infinite~$\chi $ limit was
  introduced in Ref.~\cite {rams18}}\BibitemShut {NoStop}%
\bibitem [{\citenamefont {Sandvik}\ and\ \citenamefont
  {Evertz}(2010)}]{Sandvik10}%
  \BibitemOpen
  \bibfield  {author} {\bibinfo {author} {\bibfnamefont {A.~W.}\ \bibnamefont
  {Sandvik}}\ and\ \bibinfo {author} {\bibfnamefont {H.~G.}\ \bibnamefont
  {Evertz}},\ }\href {\doibase 10.1103/PhysRevB.82.024407} {\bibfield
  {journal} {\bibinfo  {journal} {Phys. Rev. B}\ }\textbf {\bibinfo {volume}
  {82}},\ \bibinfo {pages} {024407} (\bibinfo {year} {2010})}\BibitemShut
  {NoStop}%
\bibitem [{\citenamefont {Cardy}(1986{\natexlab{a}})}]{cardy_operator_1986}%
  \BibitemOpen
  \bibfield  {author} {\bibinfo {author} {\bibfnamefont {J.~L.}\ \bibnamefont
  {Cardy}},\ }\href {\doibase 10.1016/0550-3213(86)90552-3} {\bibfield
  {journal} {\bibinfo  {journal} {Nucl. Phys. B}\ }\textbf {\bibinfo {volume}
  {270}},\ \bibinfo {pages} {186} (\bibinfo {year}
  {1986}{\natexlab{a}})}\BibitemShut {NoStop}%
\bibitem [{\citenamefont {Cardy}(1986{\natexlab{b}})}]{cardy_effect_1986}%
  \BibitemOpen
  \bibfield  {author} {\bibinfo {author} {\bibfnamefont {J.~L.}\ \bibnamefont
  {Cardy}},\ }\href {\doibase 10.1016/0550-3213(86)90596-1} {\bibfield
  {journal} {\bibinfo  {journal} {Nucl. Phys. B}\ }\textbf {\bibinfo {volume}
  {275}},\ \bibinfo {pages} {200} (\bibinfo {year}
  {1986}{\natexlab{b}})}\BibitemShut {NoStop}%
\bibitem [{\citenamefont {Rychkov}\ and\ \citenamefont
  {Tan}(2015)}]{rychkov_epsilon_2015}%
  \BibitemOpen
  \bibfield  {author} {\bibinfo {author} {\bibfnamefont {S.}~\bibnamefont
  {Rychkov}}\ and\ \bibinfo {author} {\bibfnamefont {Z.~M.}\ \bibnamefont
  {Tan}},\ }\href {\doibase 10.1088/1751-8113/48/29/29FT01} {\bibfield
  {journal} {\bibinfo  {journal} {J. Phys. A: Math. Theor.}\ }\textbf {\bibinfo
  {volume} {48}},\ \bibinfo {pages} {29FT01} (\bibinfo {year}
  {2015})}\BibitemShut {NoStop}%
\bibitem [{\citenamefont {Gliozzi}\ \emph {et~al.}(2017)\citenamefont
  {Gliozzi}, \citenamefont {Guerrieri}, \citenamefont {Petkou},\ and\
  \citenamefont {Wen}}]{gliozzi_generalized_2017}%
  \BibitemOpen
  \bibfield  {author} {\bibinfo {author} {\bibfnamefont {F.}~\bibnamefont
  {Gliozzi}}, \bibinfo {author} {\bibfnamefont {A.}~\bibnamefont {Guerrieri}},
  \bibinfo {author} {\bibfnamefont {A.~C.}\ \bibnamefont {Petkou}}, \ and\
  \bibinfo {author} {\bibfnamefont {C.}~\bibnamefont {Wen}},\ }\href {\doibase
  10.1103/PhysRevLett.118.061601} {\bibfield  {journal} {\bibinfo  {journal}
  {Phys. Rev. Lett}\ }\textbf {\bibinfo {volume} {118}} (\bibinfo {year}
  {2017}),\ 10.1103/PhysRevLett.118.061601}\BibitemShut {NoStop}%
\bibitem [{\citenamefont {Rams}\ \emph {et~al.}(2018)\citenamefont {Rams},
  \citenamefont {Czarnik},\ and\ \citenamefont {Cincio}}]{rams18}%
  \BibitemOpen
  \bibfield  {author} {\bibinfo {author} {\bibfnamefont {M.~M.}\ \bibnamefont
  {Rams}}, \bibinfo {author} {\bibfnamefont {P.}~\bibnamefont {Czarnik}}, \
  and\ \bibinfo {author} {\bibfnamefont {L.}~\bibnamefont {Cincio}},\ }\href
  {http://arxiv.org/abs/1801.08554} {\bibfield  {journal} {\bibinfo  {journal}
  {arXiv:1801.08554}\ } (\bibinfo {year} {2018})}\BibitemShut {NoStop}%
\end{thebibliography}%

\end{document}